\documentclass[aps,prd,twocolumn,showpacs,10pt,superscriptaddress,preprintnumbers,nofootinbib]{revtex4-1}
\usepackage[utf8]{inputenc}
\usepackage{bm,slashed,braket}
\usepackage[dvipsnames,table,svgnames]{xcolor}
\usepackage[normalem]{ulem}
\usepackage{colortbl}
\usepackage{multirow,boldline, booktabs}
\usepackage{makecell}
\usepackage{hhline}
\usepackage{MnSymbol}
\usepackage[colorlinks=true,
            linkcolor=blue,
            urlcolor=blue,
            citecolor=teal,          
            bookmarks=true,
            bookmarksnumbered=true,
            breaklinks=true,
            pdfpagemode=Fullscreen,
            pdfstartview=FitBH]{hyperref}
\allowdisplaybreaks[4]
\usepackage[capitalise]{cleveref}
\usepackage{orcidlink}
\usepackage{cancel}
\usepackage{tikz} 
\usepackage{tkz-euclide}
\usetikzlibrary{backgrounds} 
\usetikzlibrary{decorations.pathmorphing}
\usetikzlibrary{arrows.meta}
\usetikzlibrary{shapes.misc}
\tikzset{
mystyle/.style={line width=1, baseline, scale=0.6, every node/.style={scale=1}},
photon/.style={decorate, decoration={snake, segment length=1.5 mm, amplitude=0.5mm}, draw=black, thick},
v/.style={decorate, draw, decoration={snake, segment length=2.mm, amplitude=0.5mm}},
f/.style={draw, decoration={markings,mark=at position #1 with {\arrow[]{Latex[length=1.5mm,width=1.5mm]}}},
    postaction={decorate},node contents=#1},
f/.default=.6,
fb/.style={draw,decoration={markings,mark=at position #1 with {\arrowreversed[]{Latex[length=1.5mm,width=1.5mm]}}},
    postaction={decorate},node contents=#1},
fb/.default=.6,
s/.style={dashed,draw, postaction={decorate},
        decoration={markings,mark=at position .7 with {\arrow[very thick]{latex}}}},
sb/.style={dashed,draw, postaction={decorate},
        decoration={markings,mark=at position .55 with {\arrowreversed[draw=black,very thick]{latex}}}},
snar/.style={dashed,draw,line width =1.25pt},
gluon/.style={decorate,
 decoration={coil,amplitude=2pt, segment length=3.5pt,  pre length=.1cm, post length=.1cm}},
}

\newcommand{\tr}{\mbox{Tr}}
\newcommand{\calO}{\mathcal{O}}
\newcommand{\calM}{\mathcal{M}}
\newcommand{\N}{ {\tt N} }
\newcommand{\C}{ {\tt C} }
\newcommand{\tL}{ {\tt L} }
\newcommand{\tR}{ {\tt R} }
\newcommand{\hc}{\text{H.c.}}
\newcommand{\GeV}{\rm GeV}

\begin{document}

\title{Nucleon decays into three leptons: Noncontact contributions}

\author{Jing Chen\,\orcidlink{0009-0000-7697-6358}}
\email{cjing@m.scnu.edu.cn}
\author{Yi Liao\,\orcidlink{0000-0002-1009-5483}}
\email{liaoy@m.scnu.edu.cn}
\author{Xiao-Dong Ma\,\orcidlink{0000-0001-7207-7793}}
\email{maxid@scnu.edu.cn}
\author{Hao-Lin Wang\,\orcidlink{0000-0002-2803-5657}}
\email{whaolin@m.scnu.edu.cn}
\affiliation{State Key Laboratory of Nuclear Physics and
Technology, Institute of Quantum Matter, South China Normal
University, Guangzhou 510006, China}
\affiliation{Guangdong Basic Research Center of Excellence for
Structure and Fundamental Interactions of Matter, Guangdong
Provincial Key Laboratory of Nuclear Science, Guangzhou
510006, China}

\begin{abstract}
We investigate baryon number violating (BNV) nucleon decays into three leptons from noncontact contributions that are induced by dimension-6 (dim-6) BNV operators in low-energy effective field theory (LEFT) with an exchange of a baryon, meson, lepton, or photon field. We systematically classify all these processes that change lepton flavor by one unit and formulate their decay widths in terms of the dim-6 LEFT Wilson coefficients. By applying constraints on these Wilson coefficients derived from current experimental limits on BNV two-body nucleon decays, we obtain stringent bounds on the rates of these triple-lepton modes. These bounds vary significantly from one dim-6 operator to another under consideration. Our results for the $\Delta(B-L)=0$ modes differ by several orders of magnitude from previous phase-space estimates in the literature, thereby providing a more reliable assessment of their potential occurrence. 
In addition, we provide improved bounds on $\Delta(B+L)=0$ modes compared to the existing experimental limits. 

\end{abstract}

\maketitle 

%%%%%%%%%%%%%%%%%%%%%%%%
\section{Introduction}
%%%%%%%%%%%%%%%%%%%%%%%%

Baryon number violation (BNV) is predicted in various scenarios beyond the standard model (SM)~\cite{Pati:1974yy,Pati:1973uk,Georgi:1974sy,Weinberg:1979sa,Wilczek:1979hc,Weinberg:1980bf,Nath:2006ut}, and searching for nucleon decay offers the most practical and feasible way to test it, thanks to the abundance of nucleons in terrestrial experimental setups. Among various nucleon decay modes, exotic nucleon decays into three leptons are particularly intriguing, due to their clear experimental signatures compared to the conventional two-body modes involving a meson. Over the past few decades, experiments such as IMB~\cite{Irvine-Michigan-Brookhaven:1983iap,McGrew:1999nd}, Super-Kamiokande~\cite{Super-Kamiokande:2014pqx,Super-Kamiokande:2020tor}, and SNO+~\cite{SNO:2018ydj,SNO:2022trz} have searched for these exotic decays and established stringent limits on their occurrence, setting lower bounds on the partial lifetimes $\Gamma^{-1}$ (aka the inverse partial decay width) at the level of $10^{30}$ to $10^{34}$ years \cite{ParticleDataGroup:2024cfk}. 
The current and upcoming neutrino experiments, including JUNO~\cite{JUNO:2015zny}, Hyper-Kamiokande~\cite{Hyper-Kamiokande:2018ofw}, DUNE~\cite{DUNE:2020ypp}, and Theia~\cite{Theia:2019non}, are expected to further probe these modes with unprecedented sensitivity. 

In light of upcoming experimental efforts, it is timely to examine these decays systematically from a theoretical perspective. 
Focusing on SM leptons, the nucleon triple-lepton decay modes ${\tt N} \to l_x l_y l _z$, with ${\tt N}=p,n$ and $l_{x,y,z}=e^\pm,\mu^\pm,\nu_{e,\mu,\tau},\bar\nu_{e,\mu,\tau}$, can be classified according to a generalized lepton flavor number $\Delta F_L=1,~3$. Here, 
$\Delta F_L \equiv |\Delta F_e|+|\Delta F_\mu|+|\Delta F_\tau|$, with $\Delta F_i\, (i=e,\mu,\tau)$ denoting the number of change of the $i$-flavor. 
The $\Delta F_L=1$ class corresponds to the cases where only one lepton flavor changes by one unit, 
and these modes are listed in \cref{tab:DeltaLF1process}.
Working in the low energy effective field theory (LEFT)~\cite{Jenkins:2017jig,Liao:2020zyx},  the leading order contributions to the $\Delta F_L=1$ decays are from noncontact diagrams that are generated by dim-6 BNV interactions augmented by SM interactions with the exchange of a lepton, baryon, photon, or meson. 
In contrast, the decays with $\Delta F_L=3$ can only be mediated at the leading order by local dim-9 LEFT interactions. 

\begin{table}[t]
\centering
\resizebox{0.6\linewidth}{!}{
\begin{tabular}{|c|c|c|}
\hline
~Class~ & $\Delta(B-L) = 0$ & $\Delta(B+L) = 0$ 
\\\hline
\multirow{4}*{\rotatebox[origin=c]{90}{
Mode}}
&~~~$p \to \ell_x^- \ell_y^+ \ell_z^+ $~~~
&~~~$n \to \ell_x^+ \ell_y^- \nu_z $~~~
\\\cline{2-3}
& $n \to \ell_x^- \ell_y^+ \bar\nu_z $ 
& $p \to \ell_x^+ \nu_y \nu_z $ 
\\\cline{2-3}
& $p \to \nu_x \ell_y^+ \bar \nu_z $ 
& $n \to \bar\nu_x \nu_y \nu_z $ 
\\\cline{2-3}
& $n \to \nu_x  \bar\nu_y \bar \nu_z $  &  
\\\hline
\end{tabular}}
\caption{Classification of all possible nucleon triple-lepton decay modes with $\Delta F_L=1$. 
The lepton flavor indices $x,y,z$ are related by $x=y$ or $x=z$.}
\vspace{-1em}
\label{tab:DeltaLF1process}
\end{table}

In this Letter, we focus on the $\Delta F_L=1$ decays while leaving those with $\Delta F_L=3$ in the upcoming paper~\cite{Liao:2025wxk}. 
We aim to derive bounds on the partial lifetimes of the $\Delta F_L=1$ decays in chiral perturbation theory (ChPT), assuming that their dominant contributions arise from dim-6 LEFT BNV interactions. These bounds are obtained by utilizing available constraints on the Wilson coefficients (WCs) of the relevant dim-6 operators. Our final results, summarized in \cref{tab:ResBmL=0,tab:ResBpL=0}, show that the newly derived bounds vary from one operator to another by several orders of magnitude, and significantly improve upon the existing experimental limits. 

The remainder of this Letter is organized as follows. In \cref{sec:EFTframework}, we begin by collecting the relevant dim-6 BNV operators along with their  counterparts in ChPT. Next, in \cref{sec:LOdiagrams}, we examine the leading-order Feynman diagrams that contribute to these nucleon decay modes due to dim-6 operators. Our expressions for the decay widths, together with the newly derived bounds, are presented in \cref{tab:ResBmL=0,tab:ResBpL=0} of \cref{sec:results}. 
Our conclusions are summarized in \cref{sec:summary}.
Additionally, the involved SM four-fermion interactions and chiral interactions are collected in \cref{app:SMweak} and \cref{app:BNV-ChPT,app:BNC-ChPT}, respectively,
while the experimental bounds on the two-body nucleon decays and the branching ratios of the meson leptonic  decays are summarized in \cref{app:meson_decay}.

%%%%%%%%%%%%%%%%%%%%%%%%
\section{EFT descriptions of $\Delta F_L=1$ nucleon triple-lepton decays}
\label{sec:EFTframework}
%%%%%%%%%%%%%%%%%%%%%%%%

\begin{table}[h]
\center
\resizebox{\linewidth}{!}{
\renewcommand{\arraystretch}{1.3}
\begin{tabular}{|c|c|c|c|}
\hline
\multicolumn{2}{|c|}{$\Delta(B-L)= 0$} &
\multicolumn{2}{c|}{$\Delta(B+L)=0$}
\\
\hline
~$\calO_{\nu\rm dud}^{\tL\tL}$~  & 
~$(\overline{\nu_{\tL}^{\C}} d_{\tL}^\alpha) 
(\overline{u_{\tL}^{\beta \C}} d_{\tL}^\gamma)
\epsilon_{\alpha \beta \gamma}$~  &
~$\calO_{\bar{\ell}\rm ddd}^{\tL\tL}$~  &
~$(\overline{\ell_{\tR}} d_{\tL}^\alpha) 
(\overline{d_{\tL}^{\beta \C}} d_{\tL}^\gamma)
\epsilon_{\alpha \beta \gamma}$~
\\
\hline
$\calO_{\ell\rm udu}^{\tL\tL}$  & 
$(\overline{\ell_{\tL}^{\C}} u_{\tL}^\alpha) 
(\overline{d_{\tL}^{\beta \C}} u_{\tL}^\gamma)
\epsilon_{\alpha \beta \gamma}$  &
$\calO_{\bar{\nu}\rm dud}^{\tR\tL}$  &
$(\overline{\nu_{\tL}} d_{\tR}^\alpha) 
(\overline{u_{\tL}^{\beta \C}} d_{\tL}^\gamma)
\epsilon_{\alpha \beta \gamma}$ 
\\
\hline
$\calO_{\ell\rm duu}^{\tR\tL}$  & 
$(\overline{\ell_{\tR}^{\C}} d_{\tR}^\alpha) 
(\overline{u_{\tL}^{\beta \C}} u_{\tL}^\gamma)
\epsilon_{\alpha \beta \gamma}$  &
$\calO_{\bar{\nu}\rm udd}^{\tR\tL}$  &
$(\overline{\nu_{\tL}} u_{\tR}^\alpha) 
(\overline{d_{\tL}^{\beta \C}} d_{\tL}^\gamma)
\epsilon_{\alpha \beta \gamma}$ 
\\
\hline
$\calO_{\ell\rm udu}^{\tR\tL}$  & 
$(\overline{\ell_{\tR}^{\C}} u_{\tR}^\alpha) 
(\overline{d_{\tL}^{\beta \C}} u_{\tL}^\gamma)
\epsilon_{\alpha \beta \gamma}$  &
$\calO_{\bar{\ell}\rm ddd}^{\tR\tL}$  &
$(\overline{\ell_{\tL}} d_{\tR}^\alpha) 
(\overline{d_{\tL}^{\beta \C}} d_{\tL}^\gamma)
\epsilon_{\alpha \beta \gamma}$ 
\\
\hline
$\calO_{\ell\rm duu}^{\tL\tR}$  & 
$(\overline{\ell_{\tL}^{\C}} d_{\tL}^\alpha) 
(\overline{u_{\tR}^{\beta \C}} u_{\tR}^\gamma)
\epsilon_{\alpha \beta \gamma}$  &
$\calO_{\bar{\ell}\rm ddd}^{\tL\tR}$  &
$(\overline{\ell_{\tR}} d_{\tL}^\alpha) 
(\overline{d_{\tR}^{\beta \C}} d_{\tR}^\gamma)
\epsilon_{\alpha \beta \gamma}$ 
\\
\hline
$\calO_{\ell\rm udu}^{\tL\tR}$  &  
$(\overline{\ell_{\tL}^{\C}} u_{\tL}^\alpha) 
(\overline{d_{\tR}^{\beta \C}} u_{\tR}^\gamma)
\epsilon_{\alpha \beta \gamma}$  &
$\calO_{\bar{\nu}\rm dud}^{\tR\tR}$  &
$(\overline{\nu_{\tL}} d_{\tR}^\alpha) 
(\overline{u_{\tR}^{\beta \C}} d_{\tR}^\gamma)
\epsilon_{\alpha \beta \gamma}$ 
\\
\hline
$\calO_{\nu\rm ddu}^{\tL\tR}$  & 
$(\overline{\nu_{\tL}^{\C}} d_{\tL}^\alpha) 
(\overline{d_{\tR}^{\beta \C}} u_{\tR}^\gamma)
\epsilon_{\alpha \beta \gamma}$  &
$\calO_{\bar{\ell}\rm ddd}^{\tR\tR}$  &
$(\overline{\ell_{\tL}} d_{\tR}^\alpha) 
(\overline{d_{\tR}^{\beta \C}} d_{\tR}^\gamma)
\epsilon_{\alpha \beta \gamma}$ 
\\
\hline
$\calO_{\nu\rm udd}^{\tL\tR}$  & 
$(\overline{\nu_{\tL}^{\C}} u_{\tL}^\alpha) 
(\overline{d_{\tR}^{\beta \C}} d_{\tR}^\gamma)
\epsilon_{\alpha \beta \gamma}$  &  &
\\
\cline{1-2}
$\calO_{\ell\rm udu}^{\tR\tR}$  & 
$(\overline{\ell_{\tR}^{\C}} u_{\tR}^\alpha) 
(\overline{d_{\tR}^{\beta \C}} u_{\tR}^\gamma)
\epsilon_{\alpha \beta \gamma}$  &  &
\\
\hline
\end{tabular} }
\caption{Dim-6 BNV operators in the LEFT with $\Delta B=1$ and $\Delta L=\pm 1$. $\alpha,\beta,\gamma$ are color indices while flavor indices are omitted for simplicity. }
\label{tab:dim6ope}
\end{table}

The BNV nucleon decays are well described in the LEFT framework~\cite{Jenkins:2017jig,Liao:2020zyx,Abbott:1980zj,Claudson:1981gh,Beneito:2023xbk,Gargalionis:2024nij,Liao:2025vlj,Liao:2025sqt}, which serves as a systematic approach to processes occurring below the electroweak scale, $\Lambda_{\tt EW}$. In LEFT, the lowest-order BNV interactions appear at dimension 6 and are categorized according to their net baryon ($B$) and lepton ($L$) number changes into two classes: those with $\Delta(B-L)=0$ and those with $\Delta(B+L)=0$, as summarized in \cref{tab:dim6ope}. They consist of three quark fields and one lepton field, which can be either a charged lepton or a neutrino.

To calculate nucleon decay matrix elements due to the above dim-6 operators involving light $u,d,s$ quarks, a powerful and systematic method is to match the quark-level interactions into hadronic counterparts through ChPT, with quark degrees of freedom being traded by light hadrons. In the mesonic ChPT extended with octet baryons~\cite{Jenkins:1990jv,Bijnens:1985kj,Oller:2006yh}, the relevant hadronic degrees of freedom are the octet meson and baryon fields that are organized in matrix form,
\begin{subequations}
\begin{align}
\Sigma(x) & = \xi^2(x) = \exp\Big(\frac{i\sqrt{2}\Pi(x)}{F_0}\Big) , 
\\
\Pi(x) & =
\begin{pmatrix}
\frac{\pi^{0}}{\sqrt{2}}+\frac{\eta}{\sqrt{6}}
&\pi{+}&K^{+}
\\
\pi^{-}
&-\frac{\pi^{0}}{\sqrt{2}}+\frac{\eta}{\sqrt{6}}
&K^{0}
\\
K^{-}&\bar{K}^{0}&-\sqrt{\frac{2}{3}}\eta
\end{pmatrix},
\\
B(x)& =
\begin{pmatrix}
\frac{\Sigma^{0}}{\sqrt{2}}+\frac{\Lambda^{0}}{\sqrt{6}}&\Sigma^{+}&p
\\
\Sigma^{-}&-\frac{\Sigma^{0}}{\sqrt{2}}+\frac{\Lambda^{0}}{\sqrt{6}}&n
\\
\Xi^{-}&\Xi^{0}&-\sqrt{\frac{2}{3}}\Lambda^{0}
\end{pmatrix}.
\end{align}
\end{subequations}
Here $F_0 = (86.2 \pm 0.5)\,{\rm MeV}$ denotes the pion decay constant in the chiral limit. Under the QCD chiral group, $G_\chi=\rm SU(3)_\tL\otimes SU(3)_\tR$, the meson field $\Sigma$ transforms as $\Sigma \rightarrow \hat L \Sigma \hat R^{\dagger}$, where $(\hat L,\hat R)\in G_\chi$. $\xi$ transforms as $\xi \rightarrow \hat L \xi h^{\dagger} = \hat h \xi \hat R^{\dagger}$, where $\hat h$ is a function of $\hat L$, $\hat R$, and $\xi$, and the baryon field transforms covariantly as $B \rightarrow \hat h B \hat h^{\dagger}$.

The relevant operators responsible for nucleon decays involve light $u,d,s$ quarks and transform under $G_\chi$ as irreducible representations (irreps) $\pmb{8}_{\tL(\tR)}\otimes\pmb{1}_{\tR(\tL)}$ and $\bar{\pmb{3}}_{\tL(\tR)} \otimes \pmb{3}_{\tR(\tL)}$. 
Denoting the triple-quark component of each operator by ${\cal N}_i$ and the product of the lepton field ($\overline{\Psi}_i$) and its corresponding WC ($C_i$) as the spurion field ${\cal P}_i=C_i\overline{\Psi}_i$, the  BNV Lagrangian reads 
\begin{align}
{\cal L}_{\slashed B} = {\rm Tr}\big[\mathcal{P}_{\pmb{8}_{\tL} \otimes \pmb{1}_{\tR}}
\mathcal{N}_{\pmb{8}_{\tL} \otimes \pmb{1}_{\tR}}
+ \mathcal{P}_{\pmb{3}_\tL \otimes \bar{\pmb{3}}_\tR} \mathcal{N}_{\bar{\pmb{3}}_\tL \otimes \pmb{3}_\tR}\big] 
+\tL\leftrightarrow \tR.
\label{eq:LEFTBNVLag}
\end{align}
Here, the triple-quark components and the spurion fields have been organized as matrices in flavor space. For reference, they are collected in \cref{eq:3qpart,eq:spurion} of \cref{app:BNV-ChPT}. Under $(\hat{L},\hat{R})\in G_\chi$, they transform as
\begin{subequations}
\begin{align}
\mathcal{N}_{{\pmb{8}_{\tL} \otimes \pmb{1}_{\tR}}}
& \to \hat{L}\mathcal{N}_{{\pmb{8}_{\tL} \otimes \pmb{1}_{\tR}}} \hat{L}^\dagger, \quad
\mathcal{N}_{\bar{\pmb{3}}_\tL \otimes \pmb{3}_\tR}\to \hat{R}\mathcal{N}_{\bar{\pmb{3}}_\tL \otimes \pmb{3}_\tR} \hat{L}^\dagger,
\\
\mathcal{P}_{{\pmb{8}_{\tL} \otimes \pmb{1}_{\tR}}}
&\to \hat{L}\mathcal{P}_{{\pmb{8}_{\tL} \otimes \pmb{1}_{\tR}}} \hat{L}^\dagger, \quad
~\mathcal{P}_{\pmb{3}_\tL \otimes \bar{\pmb{3}}_\tR}\to \hat{L}\mathcal{P}_{\pmb{3}_\tL \otimes \bar{\pmb{3}}_\tR} \hat{R}^\dagger.
\end{align}
\end{subequations}
At leading chiral order, their chiral realizations take the following form~\cite{Claudson:1981gh, Fan:2024gzc,Liao:2025vlj}:
\begin{align}
{\cal L}_{\tt \slashed{B}}^{\tt ChPT}
& =
c_1 \tr\left[ 
{\cal P}_{\bar{\pmb{3}}_\tL \otimes \pmb{3}_\tR}
\xi B_{\tL}\xi
-{\cal P}_{\pmb{3}_\tL \otimes \bar{\pmb{3}}_\tR}
\xi^{\dagger} B_{\tR}\xi^{\dagger} \right]
\nonumber
\\
& +c_2 \tr \left[
{\cal P}_{\pmb{8}_{\tL} \otimes \pmb{1}_{\tR}}
\xi B_{\tL}\xi^{\dagger} -{\cal P}_{\pmb{1}_\tL \otimes \pmb{8}_\tR}
\xi^{\dagger}B_{\tR}\xi \right]+\hc,
\label{Eq.BNV-ChPT}
\end{align}
where $B_{\tL(\tR)}\equiv P_{\tL(\tR)} B$ represent the chiral baryon fields with $P_{\tR,\tL}=(1\pm\gamma_5)/2$ being the chiral projectors. The low energy constants (LECs) $c_1$ and $c_2$ are also denoted by $\alpha$ and $\beta$ in the literature, and the recent lattice QCD calculations yield $c_1=-0.01257(111)\,\GeV^3$ and $c_2=0.01269(107)\,\GeV^3$~\cite{Yoo:2021gql}. By expanding \cref{Eq.BNV-ChPT} to the desired order in the pseudoscalar meson fields, one obtains the explicit BNV interaction terms involving one lepton, one baryon, and any number of meson fields. These results are summarized in \cref{app:BNV-ChPT}.

%%%%%%%%%%%%%%%%%%%%%%%%%
\begin{table*}[t]
\centering
\resizebox{0.9\linewidth}{!}{
\renewcommand{\arraystretch}{1.}
{\large
\begin{tabular}{|c|l l|c|c|c|c|}
\hline
~~~~~~~~&\multicolumn{2}{c|}{Mode}&  \multicolumn{4}{c|}{Feynman diagrams} 
\\\hline
%%%%%
\multirow{22}{*}{\rotatebox[origin=c]{90}{$\Delta(B-L)=0$ } }
&\multicolumn{2}{c|}{$\makecell{
 p\to \ell^{-}_{x}\ell^{+}_{x} \ell^{+}_{y}\\~~~(xy=ee, \mu\mu, e\mu,\mu e)~~~}$}   &
\begin{tikzpicture}[mystyle,scale=1.0]
\begin{scope}
\draw[f] (-3.2,-0.8) node[left]{$p~$} -- (-2.0,-0.8);
\draw[f] (0.0,1.0) node[right]{$\ell^{+}_{x}~$} -- (-0.8,-0.1);
\draw[f] (-0.8,-0.1)--(0.7,0.2) node[right]{$\ell^{-}_{x}$}  (-1.7,-0.1);
\draw[photon] (-2.0,-0.8)-- (-0.8,-0.1)node[midway,xshift=-5pt,yshift=8pt]{$\gamma$};
\draw[f] (-2.0,-0.8) -- (-0.5,-0.8)node[midway,yshift = - 8 pt]{$p$};
\draw[f] (0.7,-0.8)node[right]{$\ell^{+}_{y}$} -- (-0.5,-0.8);
\draw[draw=cyan,fill=cyan](-0.5,-0.8) circle (0.13cm);
\end{scope}
\end{tikzpicture}  
& 
\begin{tikzpicture}[mystyle,scale=1.0]
\begin{scope}
\draw[f] (-3.2,-0.8) node[left]{$p~$} -- (-2.1,-0.8);
\draw[f] (0.5,1.0) node[right]{$\ell^{+}_{x}~$} -- (-0.3,-0.2);
\draw[f] (-0.3,-0.2)--(0.7,0.2) node[right]{$\ell^{-}_{x}$}  (-1.7,-0.1);
\draw[photon] (-0.3,-0.2)-- (-1.0,-0.8)node[midway,xshift=-5pt,yshift=8pt]{$\gamma$};
\draw[f] (-0.5,-0.8) -- (-2.1,-0.8)node[midway,yshift = - 8 pt]{$\ell $};
\draw[f] (0.7,-0.8)node[right]{$\ell^{+}_{y}$} -- (-0.5,-0.8);
\draw[draw=cyan,fill=cyan](-2.1,-0.8) circle (0.13cm);
\end{scope}
\end{tikzpicture} 
&
\begin{tikzpicture}[mystyle,scale=1.0]
\begin{scope}
\draw[f] (-3.2,-0.8) node[left]{$p~$} -- (-2.0,-0.8);
\draw[f] (0.0,1.0) node[right]{$\ell^+_{x}~$} -- (-0.8,-0.1);
\draw[f] (-0.8,-0.1)--(0.7,0.2) node[right]{$\ell^-_{x}$}  (-1.7,-0.1);
\draw[snar] (-2.0,-0.8)-- (-0.8,-0.1) node[midway,xshift=-12pt,yshift=10pt]{\small\colorbox{gray!15}{$\pi^0,\eta$},$K^0$};
\draw[f] (-2.0,-0.8) -- (-0.5,-0.8)node[midway,yshift = - 9 pt]{\small\colorbox{gray!15}{$p$},$\Sigma^+$};
\draw[f] (-0.5,-0.8) -- (0.7,-0.8)node[right]{$\ell^+_{y}$};
\draw[draw=cyan,fill=cyan](-0.5,-0.8) circle (0.13cm);
\draw[draw=black,fill=white] (-0.8,-0.1) circle (0.12cm);
\draw[draw=purple,fill=purple] (-2.1,-0.9) rectangle(-1.9,-0.7);
\end{scope}
\end{tikzpicture}  
&
\begin{tikzpicture}[mystyle,scale=1.0]
\begin{scope}
\draw[f] (-2.8,-0.8) node[left]{$p~$} -- (-1.6,-0.8);
\draw[f] (0.3,1.0) node[right]{$\ell^+_{x}$} -- (-0.6,-0.2);
\draw[f] (-0.6,-0.2)-- (0.5,0.15) node[right]{$\ell^-_{x}$};
\draw[snar] (-0.6,-0.2) -- (-1.6,-0.8)node[midway,xshift=-12pt,yshift=10pt]{\small\colorbox{gray!15}{$\pi^0,\eta$},$K^0$};
\draw[f] (-1.6,-0.8) -- (0.5,-0.8)node[right]{$\ell^+_{y}$};
\draw[draw=cyan,fill=cyan](-1.6,-0.8) circle (0.13cm);
\draw[draw=black,fill=white] (-0.6,-0.2) circle (0.12cm);
\end{scope}
\end{tikzpicture} 
\\%%
\cline{2-7}
& \multirow{1}{*}{\rotatebox{90}{$
\makecell{n\to\ell^{-}_{x} \ell^{+}_{x} \bar{\nu}_{x}\\(x=e,\mu)}$\hspace{0.3cm} } }
&\multicolumn{1}{|c|}{ $\makecell{
n\to \ell^{-}_{x} \ell^{+}_{x} \bar{\nu}_{y}\\
(xy=e \mu ,e \tau ,\\ 
\mu e,\mu \tau) } $ }  
&
\begin{tikzpicture}[mystyle,scale=1.0]
\begin{scope}
\draw[f] (-3.2,-0.8) node[left]{$n~$} -- (-2.0,-0.8);
\draw[f] (0.0,1.0) node[right]{$\ell^{+}_{x}~$} -- (-0.8,-0.1);
\draw[f] (-0.8,-0.1)--(0.7,0.2) node[right]{$\ell^{-}_{x}$}  (-1.7,-0.1);
\draw[photon] (-2.0,-0.8)-- (-0.8,-0.1)node[midway,xshift=-5pt,yshift=8pt]{$\gamma$};
\draw[f] (-2.0,-0.8) -- (-0.5,-0.8)node[midway,yshift = - 8 pt]{$n$};
\draw[f] (0.7,-0.8)node[right]{$\bar{\nu}_{y}$} -- (-0.5,-0.8);
\draw[draw=cyan,fill=cyan](-0.5,-0.8) circle (0.13cm);
\end{scope}
\end{tikzpicture}  
& ---
&
\begin{tikzpicture}[mystyle,scale=1.0]
\begin{scope}
\draw[f] (-3.2,-0.8) node[left]{$n~$} -- (-2.0,-0.8);
\draw[f] (0.0,1.0) node[right]{$\ell^+_{x}~$} -- (-0.8,-0.1);
\draw[f] (-0.8,-0.1)--(0.7,0.2) node[right]{$\ell^-_{x}$}  (-1.7,-0.1);
\draw[snar] (-2.0,-0.8)-- (-0.8,-0.1)node[midway,xshift=-12pt,yshift=10pt]{\small\colorbox{gray!15}{$\pi^0,\eta$},$K^0$};
\draw[f] (-2.0,-0.8) -- (-0.5,-0.8)node[midway,yshift = - 9 pt]{\small\colorbox{gray!15}{$n$},$\Lambda^0,\Sigma^0$};
\draw[f] (-0.5,-0.8) -- (0.7,-0.8)node[right]{$\bar\nu_{y}$};
\draw[draw=cyan,fill=cyan](-0.5,-0.8) circle (0.13cm);
\draw[draw=black,fill=white] (-0.8,-0.1) circle (0.12cm);
\draw[draw=purple,fill=purple] (-2.1,-0.9) rectangle(-1.9,-0.7);
\end{scope}
\end{tikzpicture}  
&
\begin{tikzpicture}[mystyle,scale=1.0]
\begin{scope}
\draw[f] (-2.8,-0.8) node[left]{$n~$} -- (-1.6,-0.8);
\draw[f] (0.3,1.0) node[right]{$\ell^+_{x}$} -- (-0.6,-0.2);
\draw[f] (-0.6,-0.2)-- (0.5,0.15) node[right]{$\ell^-_{x}$};
\draw[snar] (-0.6,-0.2) -- (-1.6,-0.8)node[midway,xshift=-12pt,yshift=10pt]{\small\colorbox{gray!15}{$\pi^0,\eta$},$K^0$};
\draw[f] (-1.6,-0.8) -- (0.5,-0.8)node[right]{$\bar\nu_{y}$};
\draw[draw=cyan,fill=cyan](-1.6,-0.8) circle (0.13cm);
\draw[draw=black,fill=white] (-0.6,-0.2) circle (0.12cm);
\end{scope}
\end{tikzpicture} 
\\\hhline{|~~-----|}%%
&
& \multicolumn{1}{|c|}{ $\makecell{
n\to \ell^{-}_{x} \ell^{+}_{y}\bar{\nu}_{x} \\ (xy=e\mu,\mu e) } $  }  
&\cellcolor{gray!15}
\begin{tikzpicture}[mystyle,scale=1.0]
\begin{scope}
\draw[f] (-3.2,-0.8) node[left]{$n~$} -- (-1.7,-0.8);
\draw[f] (-0.5,1) node[right]{$\bar{\nu}_{x}$} -- (-1.7,-0.8);
\draw[f] (-1.7,-0.8) --(0.7,0.5) node[right]{$\ell^{-}_{x}$} ;
\draw[f] (-1.7,-0.8) -- (-0.3,-0.8)node[midway,yshift = - 8 pt]{$p$};
\draw[f] (0.7,-0.8)node[right]{$\ell^{+}_{y}$} -- (-0.3,-0.8);
\draw[draw=cyan,fill=cyan](-0.4,-0.8) circle (0.13cm);
\draw[draw=black,fill=black] (-1.7,-0.8) circle (0.10cm);
\end{scope}
\end{tikzpicture}
&  
\begin{tikzpicture}[mystyle,scale=1.0]
\begin{scope}
\draw[f] (-3.2,-0.8) node[left]{$n~$} -- (-2.2,-0.8);
\draw[f] (0.2,1.0) node[right]{$\bar{\nu}_{x}$} -- (-1.1,-0.8);
\draw[f] (-1.1,-0.8) -- (0.7,0.3) node[right]{$\ell^{-}_{x}$};
\draw[f] (-1.1,-0.8) -- (-2.2,-0.8)node[midway,yshift = - 8 pt]{$\nu$};
\draw[f] (0.7,-0.8)node[right]{$\ell^{+}_{y}$} -- (-1.1,-0.8);
\draw[draw=cyan,fill=cyan](-2.2,-0.8) circle (0.13cm);
\draw[draw=black,fill=black] (-1.1,-0.8) circle (0.10cm);
\end{scope}
\end{tikzpicture} 
&
\begin{tikzpicture}[mystyle,scale=1.0]
\begin{scope}
\draw[f] (-3.2,-0.8) node[left]{$n~$} -- (-2.0,-0.8);
\draw[f] (0.0,1.0) node[right]{$\bar{\nu}_{x}~$} -- (-0.8,-0.1);
\draw[f] (-0.8,-0.1)--(0.7,0.2) node[right]{$\ell^{-}_{x}$}  (-1.7,-0.1);
\draw[snar] (-2.0,-0.8)-- (-0.8,-0.1)node[midway,xshift=-3pt,yshift=6pt]{$\pi$};
\draw[f] (-2.0,-0.8) -- (-0.5,-0.8)node[midway,yshift = - 8 pt]{$p$};
\draw[f] (0.7,-0.8)node[right]{$\ell^{+}_{y}$} -- (-0.5,-0.8);
\draw[draw=cyan,fill=cyan](-0.5,-0.8) circle (0.13cm);
\draw[draw=black,fill=black] (-0.8,-0.1) circle (0.10cm);
\draw[draw=purple,fill=purple] (-2.1,-0.9) rectangle(-1.9,-0.7);
\end{scope}
\end{tikzpicture}  
& 
\begin{tikzpicture}[mystyle,scale=1.0]
\begin{scope}
\draw[f] (-2.8,-0.8) node[left]{$n~$} -- (-1.6,-0.8);
\draw[f] (0.3,1.0) node[right]{$\bar{\nu}_{x}$} -- (-0.6,-0.2);
\draw[f] (-0.6,-0.2)-- (0.5,0.15) node[right]{$\ell^{-}_{x}$};
\draw[snar] (-0.6,-0.2) -- (-1.6,-0.8)node[midway,xshift=-3pt,yshift=6pt]{$\pi$};
\draw[f] (0.5,-0.8)node[right]{$\ell^{+}_{y}$} -- (-1.6,-0.8);
\draw[draw=cyan,fill=cyan](-1.6,-0.8) circle (0.13cm);
\draw[draw=black,fill=black] (-0.6,-0.2) circle (0.10cm);
\end{scope}
\end{tikzpicture} 
\\%%
\hhline{|~|------|}
& 
\multirow{1}{*}{\rotatebox{90}{$\makecell{p\to \ell^{+}_{x} \nu_{x} \bar{\nu}_{x} \\ (x=e,\mu)}$\hspace{0.3cm} } }
& 
\multicolumn{1}{|c|}{ $\makecell{ 
p\to \nu_{x} \ell^{+}_{y}  \bar{\nu}_{x} \\ (xy=\mu e,\tau e,\\ e\mu,\tau\mu)}$ }  
& 
\begin{tikzpicture}[mystyle,scale=1.0]
\begin{scope}
\draw[f] (-3.2,-0.8) node[left]{$p~$} -- (-1.7,-0.8);
\draw[f] (-0.5,1) node[right]{$\bar{\nu}_{x}~$} -- (-1.7,-0.8);
\draw[f] (-1.7,-0.8) --(0.7,0.5) node[right]{$\nu_{x}$} ;
\draw[f] (-1.7,-0.8) -- (-0.3,-0.8)node[midway,yshift = - 8 pt]{$p$};
\draw[f] (0.7,-0.8)node[right]{$\ell^{+}_{y}$} -- (-0.3,-0.8);
\draw[draw=cyan,fill=cyan](-0.4,-0.8) circle (0.13cm);
\draw[draw=black,fill=white] (-1.7,-0.8) circle (0.12cm);
\end{scope}
\end{tikzpicture}
&
\begin{tikzpicture}[mystyle,scale=1.0]
\begin{scope}
\draw[f] (-3.2,-0.8) node[left]{$p~$} -- (-2.2,-0.8);
\draw[f] (0.2,1.0) node[right]{$\bar{\nu}_{x}~$} -- (-1.1,-0.8);
\draw[f] (-1.1,-0.8) -- (0.7,0.3) node[right]{$\nu_{x}$};
\draw[f] (-1.1,-0.8) -- (-2.2,-0.8)node[midway,yshift = - 8 pt]{$\ell$};
\draw[f] (0.7,-0.8)node[right]{$\ell^{+}_{y}$} -- (-1.1,-0.8);
\draw[draw=cyan,fill=cyan](-2.2,-0.8) circle (0.13cm);
\draw[draw=black,fill=white] (-1.1,-0.8) circle (0.12cm);
\end{scope}
\end{tikzpicture}
&
\cellcolor{gray!15}
\begin{tikzpicture}[mystyle,scale=1.0]
\begin{scope}
\draw[f] (-3.2,-0.8) node[left]{$p~$} -- (-2.0,-0.8);
\draw[f] (0.0,1.0) node[right]{$\bar{\nu}_{x}~$} -- (-0.8,-0.1);
\draw[f] (-0.8,-0.1)--(0.7,0.2) node[right]{$\nu_{x}$}  (-1.7,-0.1);
\draw[snar] (-2.0,-0.8)-- (-0.8,-0.1);
\draw[f] (-2.0,-0.8) -- (-0.5,-0.8);
\draw[f] (0.7,-0.8)node[right]{$\ell^{+}_{y}$} -- (-0.5,-0.8);
\draw[draw=cyan,fill=cyan](-0.5,-0.8) circle (0.13cm);
\draw[draw=black,fill=white] (-0.8,-0.1) circle (0.12cm);
\draw[draw=purple,fill=purple] (-2.1,-0.9) rectangle(-1.9,-0.7);
\end{scope}
\end{tikzpicture}  
&
\cellcolor{gray!15}
\begin{tikzpicture}[mystyle,scale=1.0]
\begin{scope}
\draw[f] (-2.8,-0.8) node[left]{$p~$} -- (-1.6,-0.8);
\draw[f] (0.3,1.0) node[right]{$\bar{\nu}_{x}~$} -- (-0.6,-0.2);
\draw[f] (-0.6,-0.2)-- (0.5,0.15) node[right]{$\nu_{x}$};
\draw[snar] (-0.6,-0.2) -- (-1.6,-0.8);
\draw[f] (0.5,-0.8)node[right]{$\ell^{+}_{y}$} -- (-1.6,-0.8);
\draw[draw=cyan,fill=cyan](-1.6,-0.8) circle (0.13cm);
\draw[draw=black,fill=white] (-0.6,-0.2) circle (0.12cm);
\end{scope}
\end{tikzpicture}
\\%%
\hhline{|~~|-----|}
& & \multicolumn{1}{|c|}{ $\makecell{
p\to \nu_{x} \ell^{+}_{x} \bar{\nu}_{y}\\ (xy=e\mu,e\tau,\\\mu e,\mu\tau) } $ }
&\cellcolor{gray!15}
\begin{tikzpicture}[mystyle,scale=1.0]
\begin{scope}
\draw[f] (-3.2,-0.8) node[left]{$p~$} -- (-1.7,-0.8);
\draw[f] (-0.5,1) node[right]{$\ell^{+}_{x}$} -- (-1.7,-0.8);
\draw[f] (-1.7,-0.8) --(0.7,0.5) node[right]{$\nu_{x}$} ;
\draw[f] (-1.7,-0.8) -- (-0.3,-0.8)node[midway,yshift = - 8 pt]{\small$n,\Lambda^0,\Sigma^0$};
\draw[f] (0.7,-0.8)node[right]{$\bar{\nu}_{y}$} -- (-0.3,-0.8);
\draw[draw=cyan,fill=cyan](-0.4,-0.8) circle (0.13cm);
\draw[draw=black,fill=black] (-1.7,-0.8) circle (0.10cm);
\end{scope}
\end{tikzpicture} 
& 
\begin{tikzpicture}[mystyle,scale=1.0]
\begin{scope}
\draw[f] (-3.2,-0.8) node[left]{$p~$} -- (-2.2,-0.8);
\draw[f] (0.2,1.0) node[right]{$\ell^{+}_{x}$} -- (-1.1,-0.8);
\draw[f] (-1.1,-0.8) -- (0.7,0.3) node[right]{$\nu_{x}$};
\draw[f] (-1.1,-0.8) -- (-2.2,-0.8)node[midway,yshift = - 8 pt]{$\ell$};
\draw[f] (0.7,-0.8)node[right]{$\bar{\nu}_{y}$} -- (-1.1,-0.8);
\draw[draw=cyan,fill=cyan](-2.2,-0.8) circle (0.13cm);
\draw[draw=black,fill=black] (-1.1,-0.8) circle (0.10cm);
\end{scope}
\end{tikzpicture}
&
\begin{tikzpicture}[mystyle,scale=1.0]
\begin{scope}
\draw[f] (-3.2,-0.8) node[left]{$p~$} -- (-2.0,-0.8);
\draw[f] (0.0,1.0) node[right]{$\ell^{+}_{x}$} -- (-0.8,-0.1);
\draw[f] (-0.8,-0.1)--(0.7,0.2) node[right]{$\nu_{x}$}  (-1.7,-0.1);
\draw[snar] (-2.0,-0.8)-- (-0.8,-0.1)node[midway,xshift=-5pt,yshift=8pt]{\small$\pi,K$};
\draw[f] (-2.0,-0.8) -- (-0.5,-0.8)node[midway,yshift = - 8 pt]{\small$n,\Lambda^0,\Sigma^0$};
\draw[f] (0.7,-0.8)node[right]{$\bar{\nu}_{y}$} -- (-0.5,-0.8);
\draw[draw=cyan,fill=cyan](-0.5,-0.8) circle (0.13cm);
\draw[draw=black,fill=black] (-0.8,-0.1) circle (0.10cm);
\draw[draw=purple,fill=purple] (-2.1,-0.9) rectangle(-1.9,-0.7);
\end{scope}
\end{tikzpicture}
&
\begin{tikzpicture}[mystyle,scale=1.0]
\begin{scope}
\draw[f] (-2.8,-0.8) node[left]{$p~$} -- (-1.6,-0.8);
\draw[f] (0.3,1.0) node[right]{$\ell^{+}_{x}$} -- (-0.6,-0.2);
\draw[f] (-0.6,-0.2)-- (0.5,0.15) node[right]{$\nu_{x}$};
\draw[snar] (-0.6,-0.2) -- (-1.6,-0.8)node[midway,xshift=-5pt,yshift=8pt]{\small$\pi,K$};
\draw[f] (0.5,-0.8)node[right]{$\bar{\nu}_{y}$} -- (-1.6,-0.8);
\draw[draw=cyan,fill=cyan](-1.6,-0.8) circle (0.13cm);
\draw[draw=black,fill=black] (-0.6,-0.2) circle (0.10cm);
\end{scope}
\end{tikzpicture} 
\\%%
 \hhline{|~|------|}
&\multicolumn{2}{c|}{ $\makecell{ n \to \nu_{x} \bar{\nu}_{x}\bar{\nu}_{y}
\\(x,y=e,\mu,\tau)} $} &
\begin{tikzpicture}[mystyle,scale=1.0]
\begin{scope}
\draw[f] (-3.2,-0.8) node[left]{$n~$} -- (-1.7,-0.8);
\draw[f] (-1.7,-0.8) -- (0.7,0.5) node[right]{$\nu_{x}$};
\draw[f] (-0.5,1) node[right]{$\bar\nu_{x}~$} -- (-1.7,-0.8);
\draw[f] (-1.7,-0.8) -- (-0.3,-0.8)node[midway,yshift = - 8 pt]{$n$};
\draw[f] (0.7,-0.8)node[right]{$\bar\nu_{y}$} -- (-0.3,-0.8);
\draw[draw=cyan,fill=cyan](-0.4,-0.8) circle (0.13cm);
\draw[draw=black,fill=white] (-1.7,-0.8) circle (0.12cm);
\end{scope}
\end{tikzpicture}   &
\begin{tikzpicture}[mystyle,scale=1.0]
\begin{scope}
\draw[f] (-3.2,-0.8) node[left]{$n~$} -- (-2.2,-0.8);
\draw[f] (0.2,1.0) node[right]{$\bar\nu_{x}~$} -- (-1.1,-0.8);
\draw[f] (-1.1,-0.8) -- (0.7,0.3) node[right]{$\nu_{x}$};
\draw[f] (-1.1,-0.8) -- (-2.2,-0.8)node[midway,yshift = - 8 pt]{$\nu$};
\draw[f] (0.7,-0.8)node[right]{$\bar\nu_{y}$} -- (-1.1,-0.8);
\draw[draw=cyan,fill=cyan](-2.2,-0.8) circle (0.13cm);
\draw[draw=black,fill=white] (-1.1,-0.8) circle (0.12cm);
\end{scope}
\end{tikzpicture} 
&
\cellcolor{gray!15}
\begin{tikzpicture}[mystyle,scale=1.0]
\begin{scope}
\draw[f] (-3.2,-0.8) node[left]{$n~$} -- (-2.0,-0.8);
\draw[f] (0.0,1.0) node[right]{$\bar\nu_{x}~$} -- (-0.8,-0.1);
\draw[f] (-0.8,-0.1)--(0.7,0.2) node[right]{$\nu_{x}$}  (-1.7,-0.1);
\draw[snar] (-2.0,-0.8)-- (-0.8,-0.1);
\draw[f] (-2.0,-0.8) -- (-0.5,-0.8);
\draw[f] (0.7,-0.8)node[right]{$\bar\nu_{y}$} -- (-0.5,-0.8);
\draw[draw=cyan,fill=cyan](-0.5,-0.8) circle (0.13cm);
\draw[draw=black,fill=white] (-0.8,-0.1) circle (0.12cm);
\draw[draw=purple,fill=purple] (-2.1,-0.9) rectangle(-1.9,-0.7);
\end{scope}
\end{tikzpicture}  
&
\cellcolor{gray!15}
\begin{tikzpicture}[mystyle,scale=1.0]
\begin{scope}
\draw[f] (-2.8,-0.8) node[left]{$n~$} -- (-1.6,-0.8);
\draw[f] (0.3,1.0) node[right]{$\bar\nu_{x}~$} -- (-0.6,-0.2);
\draw[f] (-0.6,-0.2)-- (0.5,0.15) node[right]{$\nu_{x}$};
\draw[snar] (-0.6,-0.2) -- (-1.6,-0.8);
\draw[f] (0.5,-0.8)node[right]{$\bar\nu_{y}$} -- (-1.6,-0.8);
\draw[draw=cyan,fill=cyan](-1.6,-0.8) circle (0.13cm);
\draw[draw=black,fill=white] (-0.6,-0.2) circle (0.12cm);
\end{scope}
\end{tikzpicture}
\\\hline%%%%%
\multirow{13}{*}{\rotatebox[origin=c]{90}{$\Delta(B+L)=0$} }
& \multirow[c]{1}{*}{\rotatebox{90}{$
\makecell{n\to\ell^{-}_{x} \ell^{+}_{x}\nu_{x}\\(x=e,\mu)}$\hspace{0.3cm} } }
&\multicolumn{1}{|c|}{ $\makecell{
n\to \ell^{-}_{x} \ell^{+}_{x}  \nu_{y} 
\\ (xy=e\mu, e \tau,\\ \mu e, \mu\tau) } $}   &
\begin{tikzpicture}[mystyle,scale=1.0]
\begin{scope}
\draw[f] (-3.2,-0.8) node[left]{$n~$} -- (-2.0,-0.8);
\draw[f] (0.0,1.0) node[right]{$\ell^{+}_{x}~$} -- (-0.8,-0.1);
\draw[f] (-0.8,-0.1)--(0.7,0.2) node[right]{$\ell^{-}_{x}$}  (-1.7,-0.1);
\draw[photon] (-2.0,-0.8)-- (-0.8,-0.1)node[midway,xshift=-5pt,yshift=8pt]{$\gamma$};
\draw[f] (-2.0,-0.8) -- (-0.5,-0.8)node[midway,yshift = - 8 pt]{$n$};
\draw[f] (-0.5,-0.8) -- (0.7,-0.8)node[right]{$\nu_{y}$};
\draw[draw=cyan,fill=cyan]  (-0.5,-0.8) circle (0.13cm);
\end{scope}
\end{tikzpicture}  
& ---
& 
\begin{tikzpicture}[mystyle,scale=1.0]
\begin{scope}
\draw[f] (-3.2,-0.8) node[left]{$n~$} -- (-2.0,-0.8);
\draw[f] (0.0,1.0) node[right]{$\ell^+_{x}~$} -- (-0.8,-0.1);
\draw[f] (-0.8,-0.1)--(0.7,0.2) node[right]{$\ell^-_{x}$}  (-1.7,-0.1);
\draw[snar] (-2.0,-0.8)-- (-0.8,-0.1) node[midway,xshift=-12pt,yshift=11pt]{\small\colorbox{gray!15}{$\pi^0,\eta$},$K^0$};
\draw[f] (-2.0,-0.8) -- (-0.5,-0.8) node[midway,yshift = - 8 pt]{\small\colorbox{gray!15}{$n$},$\Lambda^0,\Sigma^0$};
\draw[f] (-0.5,-0.8) -- (0.7,-0.8)node[right]{$\nu_{y}$};
\draw[draw=cyan,fill=cyan](-0.5,-0.8) circle (0.13cm);
\draw[draw=black,fill=white] (-0.8,-0.1) circle (0.12cm);
\draw[draw=purple,fill=purple] (-2.1,-0.9) rectangle(-1.9,-0.7);
\end{scope}
\end{tikzpicture}  
&
\begin{tikzpicture}[mystyle,scale=1.0]
\begin{scope}
\draw[f] (-2.8,-0.8) node[left]{$n~$} -- (-1.6,-0.8);
\draw[f] (0.3,1.0) node[right]{$\ell^+_{x}$} -- (-0.6,-0.2);
\draw[f] (-0.6,-0.2)-- (0.5,0.15) node[right]{$\ell^-_{x}$};
\draw[snar] (-0.6,-0.2) -- (-1.6,-0.8) node[midway,xshift=-12pt,yshift=10pt]{\small\colorbox{gray!15}{$\pi^0,\eta$},$K^0$};
\draw[f] (-1.6,-0.8) -- (0.5,-0.8)node[right]{$\nu_{y}$};
\draw[draw=cyan,fill=cyan](-1.6,-0.8) circle (0.13cm);
\draw[draw=black,fill=white] (-0.6,-0.2) circle (0.12cm);
\end{scope}
\end{tikzpicture} 
\\\hhline{|~~-----|}%%%%%
& &\multicolumn{1}{|c|}{ $\makecell{
n\to \ell^{+}_{x} \ell^{-}_{y}\nu_{x} \\ (xy=e\mu,\mu e) } $}    
&\cellcolor{gray!15}  
\begin{tikzpicture}[mystyle,scale=1.0]
\begin{scope}
\draw[f] (-3.2,-0.8) node[left]{$n~$} -- (-1.7,-0.8);
\draw[f] (-0.5,1) node[right]{$\ell^{+}_{x}$} -- (-1.7,-0.8);
\draw[f] (-1.7,-0.8) --(0.7,0.5) node[right]{$\nu_{x}$} ;
\draw[f] (-1.7,-0.8) -- (-0.3,-0.8)node[midway,yshift = - 8 pt]{\small$\Sigma^-$};
\draw[f] (-0.3,-0.8) -- (0.7,-0.8)node[right]{$\ell^{-}_{y}$};
\draw[draw=cyan,fill=cyan] (-0.4,-0.8) circle (0.13cm);
\draw[draw=black,fill=black] (-1.7,-0.8) circle (0.10cm);
\end{scope}
\end{tikzpicture} 
& 
\begin{tikzpicture}[mystyle,scale=1.0]
\begin{scope}
\draw[f] (-3.2,-0.8) node[left]{$n~$} -- (-2.2,-0.8);
\draw[f] (0.2,1.0) node[right]{$\ell^{+}_{x}$} -- (-1.1,-0.8);
\draw[f] (-1.1,-0.8) -- (0.7,0.3) node[right]{$\nu_{x}$};
\draw[f] (-2.2,-0.8) -- (-1.1,-0.8)node[midway,yshift = - 8 pt]{$\nu$};
\draw[f] (-1.1,-0.8) -- (0.7,-0.8)node[right]{$\ell^{-}_{y}$};
\draw[draw=cyan,fill=cyan](-2.2,-0.8) circle (0.13cm);
\draw[draw=black,fill=black] (-1.1,-0.8) circle (0.10cm);
\end{scope}
\end{tikzpicture}   
&
\begin{tikzpicture}[mystyle,scale=1.0]
\begin{scope}
\draw[f] (-3.2,-0.8) node[left]{$n~$} -- (-2.0,-0.8);
\draw[f] (0.0,1.0) node[right]{$\ell^{+}_{x}$} -- (-0.8,-0.1);
\draw[f] (-0.8,-0.1)--(0.7,0.2) node[right]{$\nu_{x}$}  (-1.7,-0.1);
\draw[snar] (-2.0,-0.8)-- (-0.8,-0.1)node[midway,xshift=-5pt,yshift=8pt]{\small$K$};
\draw[f] (-2.0,-0.8) -- (-0.5,-0.8)node[midway,yshift = - 8 pt]{\small$\Sigma^-$};
\draw[f] (-0.5,-0.8) -- (0.7,-0.8)node[right]{$\ell^{-}_{y}$};
\draw[draw=cyan,fill=cyan](-0.5,-0.8) circle (0.13cm);
\draw[draw=black,fill=black] (-0.8,-0.1) circle (0.10cm);
\draw[draw=purple,fill=purple] (-2.1,-0.9) rectangle(-1.9,-0.7);
\end{scope}
\end{tikzpicture}  
&
\begin{tikzpicture}[mystyle,scale=1.0]
\begin{scope}
\draw[f] (-2.8,-0.8) node[left]{$n~$} -- (-1.6,-0.8);
\draw[f] (0.3,1.0) node[right]{$\ell^{+}_{x}$} -- (-0.6,-0.2);
\draw[f] (-0.6,-0.2)-- (0.5,0.15) node[right]{$\nu_{x}$};
\draw[snar] (-0.6,-0.2) -- (-1.6,-0.8)node[midway,xshift=-5pt,yshift=8pt]{\small$K$};
\draw[f] (-1.6,-0.8) -- (0.5,-0.8)node[right]{$\ell^{-}_{y}$};
\draw[draw=cyan,fill=cyan](-1.6,-0.8) circle (0.13cm);
\draw[draw=black,fill=black] (-0.6,-0.2) circle (0.10cm);
\end{scope}
\end{tikzpicture} 
\\
\hhline{|~|------|}
&\multicolumn{2}{c|}{ $\makecell{
p\to \ell^{+}_{x} \nu_{x} \nu_{y}\\ 
(xy=ee,\mu\mu, \\ e\mu,e\tau,\mu e,\mu\tau)}$ } 
&\cellcolor{gray!15} 
\begin{tikzpicture}[mystyle,scale=1.0]
\begin{scope}
\draw[f] (-3.2,-0.8) node[left]{$p~$} -- (-1.7,-0.8);
\draw[f] (-0.5,1) node[right]{$\ell^{+}_{x}$} -- (-1.7,-0.8);
\draw[f] (-1.7,-0.8) --(0.7,0.5) node[right]{$\nu_{x}$} ;
\draw[f] (-1.7,-0.8) -- (-0.3,-0.8)node[midway,yshift = - 8 pt]{\small$n,\Lambda^0,\Sigma^0$};
\draw[f] (-0.3,-0.8) -- (0.7,-0.8)node[right]{$\nu_{y}$};
\draw[draw=cyan,fill=cyan](-0.4,-0.8) circle (0.13cm);
\draw[draw=black,fill=black] (-1.7,-0.8) circle (0.10cm);
\end{scope}
\end{tikzpicture}
& ---
& 
\begin{tikzpicture}[mystyle,scale=1.0]
\begin{scope}
\draw[f] (-3.2,-0.8) node[left]{$p~$} -- (-2.0,-0.8);
\draw[f] (0.0,1.0) node[right]{$\ell^{+}_{x}$} -- (-0.8,-0.1);
\draw[f] (-0.8,-0.1)--(0.7,0.2) node[right]{$\nu_{x}$}  (-1.7,-0.1);
\draw[snar] (-2.0,-0.8)-- (-0.8,-0.1)node[midway,xshift=-5pt,yshift=8pt]{\small$\pi,K$};
\draw[f] (-2.0,-0.8) -- (-0.5,-0.8)node[midway,yshift = - 8 pt]{\small$n,\Lambda^0,\Sigma^0$};
\draw[f] (-0.5,-0.8) -- (0.7,-0.8)node[right]{$\nu_{y}$};
\draw[draw=cyan,fill=cyan](-0.5,-0.8) circle (0.13cm);
\draw[draw=black,fill=black] (-0.8,-0.1) circle (0.10cm);
\draw[draw=purple,fill=purple] (-2.1,-0.9) rectangle(-1.9,-0.7);
\end{scope}
\end{tikzpicture}  
&
\begin{tikzpicture}[mystyle,scale=1.0]
\begin{scope}
\draw[f] (-2.8,-0.8) node[left]{$p~$} -- (-1.6,-0.8);
\draw[f] (0.3,1.0) node[right]{$\ell^{+}_{x}$} -- (-0.6,-0.2);
\draw[f] (-0.6,-0.2)-- (0.5,0.15) node[right]{$\nu_{x}$};
\draw[snar] (-0.6,-0.2) -- (-1.6,-0.8)node[midway,xshift=-5pt,yshift=8pt]{\small$\pi,K$};
\draw[f] (-1.6,-0.8) -- (0.5,-0.8)node[right]{$\nu_{y}$};
\draw[draw=cyan,fill=cyan](-1.6,-0.8) circle (0.13cm);
\draw[draw=black,fill=black] (-0.6,-0.2) circle (0.10cm);
\end{scope}
\end{tikzpicture}
\\
\hhline{|~|------|}
& \multicolumn{2}{c|}{ $\makecell{
n \to \bar{\nu}_{x} \nu_{x} \nu_{y} 
\\ (x/y=e,\mu,\tau)} $} &
\begin{tikzpicture}[mystyle,scale=1.0]
\begin{scope}
\draw[f] (-3.2,-0.8) node[left]{$n~$} -- (-1.7,-0.8);
\draw[f] (-0.5,1) node[right]{$\bar\nu_{x}~$} -- (-1.7,-0.8);
\draw[f] (-1.7,-0.8) --(0.7,0.5) node[right]{$\nu_{x}$} ;
\draw[f] (-1.7,-0.8) -- (-0.3,-0.8)node[midway,yshift = - 8 pt]{$n$};
\draw[f] (-0.3,-0.8) -- (0.7,-0.8)node[right]{$\nu_{y}$};
\draw[draw=cyan,fill=cyan](-0.4,-0.8) circle (0.13cm);
\draw[draw=black,fill=white] (-1.7,-0.8) circle (0.12cm);
\end{scope}
\end{tikzpicture}   
&
\begin{tikzpicture}[mystyle,scale=1.0]
\begin{scope}
\draw[f] (-3.2,-0.8) node[left]{$n~$} -- (-2.2,-0.8);
\draw[f] (0.2,1.0) node[right]{$\bar\nu_{x}~$} -- (-1.1,-0.8);
\draw[f] (-1.1,-0.8) -- (0.7,0.3) node[right]{$\nu_{x}$};
\draw[f] (-2.2,-0.8) -- (-1.1,-0.8)node[midway,yshift = - 8 pt]{$\nu$};
\draw[f] (-1.1,-0.8) -- (0.7,-0.8)node[right]{$\nu_{y}$};
\draw[draw=cyan,fill=cyan](-2.2,-0.8) circle (0.13cm);
\draw[draw=black,fill=white] (-1.1,-0.8) circle (0.12cm);
\end{scope}
\end{tikzpicture} 
&
\cellcolor{gray!15}
\begin{tikzpicture}[mystyle,scale=1.0]
\begin{scope}
\draw[f] (-3.2,-0.8) node[left]{$n~$} -- (-2.0,-0.8);
\draw[f] (0.0,1.0) node[right]{$\bar\nu_{x}~$} -- (-0.8,-0.1);
\draw[f] (-0.8,-0.1)--(0.7,0.2) node[right]{$\nu_{x}$}  (-1.7,-0.1);
\draw[snar] (-2.0,-0.8)-- (-0.8,-0.1);
\draw[f] (-2.0,-0.8) -- (-0.5,-0.8);
\draw[f] (-0.5,-0.8) -- (0.7,-0.8)node[right]{$\nu_{y}$};
\draw[draw=cyan,fill=cyan](-0.5,-0.8) circle (0.13cm);
\draw[draw=black,fill=white] (-0.8,-0.1) circle (0.12cm);
\draw[draw=purple,fill=purple] (-2.1,-0.9) rectangle(-1.9,-0.7);
\end{scope}
\end{tikzpicture}  
&\cellcolor{gray!15}
\begin{tikzpicture}[mystyle,scale=1.0]
\begin{scope}
\draw[f] (-2.8,-0.8) node[left]{$n~$} -- (-1.6,-0.8);
\draw[f] (0.3,1.0) node[right]{$\bar\nu_{x}$} -- (-0.6,-0.2);
\draw[f] (-0.6,-0.2)-- (0.5,0.15) node[right]{$\nu_{x}$};
\draw[snar] (-0.6,-0.2) -- (-1.6,-0.8);
\draw[f] (-1.6,-0.8) -- (0.5,-0.8)node[right]{$\nu_{y}$};
\draw[draw=cyan,fill=cyan](-1.6,-0.8) circle (0.13cm);
\draw[draw=black,fill=white] (-0.6,-0.2) circle (0.12cm);
\end{scope}
\end{tikzpicture} 
\\
\hline
\end{tabular}}}
\caption{Diagrams showing leading-order noncontact contributions to nucleon triple-lepton decays. The cyan blobs denote the BNV vertices,
while the purple squares and black (hollow) dots indicate the SM strong and weak charged- (neutral-)current vertices, respectively.
}
\label{tab:LO-diagrams}
\end{table*}

Since we are concerned with noncontact contributions to nucleon decays into three leptons, additional baryon number conserving (BNC) interactions from the SM must be incorporated. These interactions include the QED interactions for charged leptons and nucleons, the strong interactions between the octet baryons and mesons, and the charged- and neutral-current weak interactions between lepton pairs and mesons or baryons. A detailed description of all these interactions can be found in \cref{app:BNC-ChPT}.

%%%%%%%%%%%%%%%%%%%%%%%%%%%%%%%%%%%%%%
\section{Leading-order diagrams}
\label{sec:LOdiagrams}
%%%%%%%%%%%%%%%%%%%%%%%%%%%%%%%%%%%%%%

Using the two- and three-point BNV vertices provided in \cref{eq:ChiLBNV-Bl,eq:ChiLBNV-BlM}, the QED vertices in \cref{eq:qed}, the strong interactions of the octet mesons and baryons in \cref{eq:qcd}, as well as the neutral- and charged-current weak interactions involving mesons and baryons in \cref{eq:mesonCC,eq:baryonNC,eq:baryonCC}, we can construct all possible leading-order Feynman diagrams for each process. 
\cref{tab:LO-diagrams} summarizes the diagrams induced by the dim-6 BNV operators for the seven classes of $\Delta F_L=1$ processes listed in \cref{tab:DeltaLF1process}. The crossed diagrams for identical final-state particles are not shown for simplicity.
These diagrams can be categorized into three types: photon-mediated diagrams, pseudoscalar-meson-mediated diagrams (shown in the last two columns), and diagrams involving weak four-fermion interactions.
The meson-mediated diagrams with a gray background vanish in the limit of $m_\nu\to 0$, while the diagrams involving weak charged-current four-fermion vertices shown in gray, as well as the meson-mediated diagrams with their intermediate particles highlighted in gray, are subdominant compared to the other retained diagrams in the presence of the same dim-6 BNV vertices. 
Therefore, we omit the diagrams related to gray in the calculations and only keep the dominant contributions from each relevant operator in every process.
Nucleon decays induced by weak interactions have also been previously  employed to constrain BNV operators involving higher generation fermions \cite{Hou:2005iu,Beneke:2024hox}.

It should be noticed that diagrams mediated by pseudoscalar mesons can also be replaced by octet vector mesons such as $\omega,\rho, K^*$. However, for a given dim-6 BNV operator, these contributions are suppressed compared to other diagrams involving the same operator shown in \cref{tab:LO-diagrams}.   
This suppression arises mainly from three factors.
First, vector mesons have relatively larger masses and widths, which leads to suppression through propagator effects. 
Second, their couplings to charged lepton currents are extremely small, as evidenced by the much smaller branching ratios of their leptonic decays, as shown in the last two columns of \cref{tab:data} in \cref{app:meson_decay}.
Third, the $\rho^0$- and $\omega$-mediated diagrams involve the same BNV operators as the photon-mediated diagrams, and we will demonstrate in the next section that the former contributions are subdominant compared to the latter.

In the following, we provide a detailed discussion of the leading contributions to the processes in each class, along with the corresponding BNV operators that induce them. We start with the $\Delta(B-L)=0$ sector.   

\pmb{$p \to \ell^{-} \ell^{+} \ell^{+}$ class:} For proton decays into three charged leptons, there are four $\Delta F_L=1$ processes: $p \to e^- e^+ e^+$, $p \to \mu^- \mu^+ \mu^+$, $p \to e^- e^+ \mu^+$, and $p \to \mu^- \mu^+ e^+$. 
They are induced by the dim-6 operators $\calO_{\ell uud}^{{\tt XY},e(\mu)}$ via photon-mediated and $\pi^0,\eta$-mediated diagrams, and by the operators $\calO_{\ell usu}^{{\tt XY},e(\mu)}$ through $K^0$-mediated diagrams.
The generic chirality labels $\tt X,Y = \tL,\tR$ are used throughout our paper. 
Among these contributions, those mediated by $\pi^0$ and $\eta$ are suppressed relative to the photon-mediated ones. 
Contributions from $K^0$-mediated diagrams are kept because they involve distinct operators. 

\pmb{$n \to \ell^{-} \ell^{+} \bar{\nu}$ class:} These processes can be divided into three groups based on their lepton flavor combinations: $n\to\ell^-_x\ell^+_x\bar\nu_{y\neq x}$, $n\to\ell^-_x\ell^+_{y\neq x}\bar\nu_x$, and $n\to\ell^-_x\ell^+_x\bar\nu_x$.  
For the first group, $n\to\ell^-_x\ell^+_x\bar\nu_{y\neq x}$, 
the dominant contributions arise from the photon-mediated diagrams involving the operators $\calO_{\nu dud}^{\tL\tR(\tL\tL),y}$, as well as from the $K^0$-mediated diagrams associated with the operators $\calO_{\nu uds}^{\tL\tR,y},\calO_{\nu dsu}^{\tL\tR(\tL\tL),y},\calO_{\nu sud}^{\tL\tR(\tL\tL),y}$. 
In the second group, $n\to\ell^-_x\ell^+_{y\neq x}\bar\nu_x$, 
the leading contributions come from the diagrams with a four-lepton vertex via the operators $\calO_{\nu dud}^{\tL\tR(\tL\tL),y}$ as well as from the $\pi^\pm$-mediated diagrams due to the operators $\calO_{\ell uud}^{{\tt XY},y}$. 
Finally, for the third group,  
which includes the processes $n \to e^{-} e^{+} \bar{\nu}_e$ and $n\to \mu^{-} \mu^{+} \bar{\nu}_\mu$, the dominant contributions are a combination of those described in the first two groups.

\pmb{$p \to \ell^{+} \nu \bar{\nu}$ class:}
Similarly to the $n \to \ell^{-} \ell^{+} \bar{\nu}$ class, the processes in this class are categorized into three groups: $p\to\nu_x\ell^+_{y\neq x}\bar\nu_x$, $p\to\nu_x\ell_x^+\bar\nu_{y\neq x}$, and $p\to\nu_x\ell_x^+\bar\nu_x$. 
In the first group, the four relevant processes ($p\to e^+\nu_\mu \bar\nu_\mu$, $p\to e^+\nu_\tau \bar\nu_\tau$, $p\to\mu^+\nu_e \bar\nu_e$, $p\to\mu^+\nu_\tau \bar\nu_\tau$) are uniquely induced by
the operators $\calO_{\ell uud}^{{\tt XY},y}$. 
For the second group, the processes can be generated both by the operators $\calO_{\ell uud}^{{\tt XY},y}$ through the four-lepton vertex diagrams and by the operators $\calO_{\nu uds}^{\tL\tR,y},\calO_{\nu dsu}^{\tL\tR(\tL\tL),y},\calO_{\nu sud}^{\tL\tR(\tL\tL),y}$($\calO_{\nu dud}^{\tL\tR(\tL\tL),y}$) via the $K^\pm(\pi^\pm)$-mediated diagrams. 
Finally, the two processes in the third group, $p \to e^+\nu_e \bar\nu_e$ and $p \to \mu^+\nu_\mu \bar\nu_\mu$, receive combined contributions from both the first and second groups.

\pmb{$n \to \nu \bar{\nu} \bar{\nu}$ class:} The processes in this group are uniquely induced by the operators $\calO_{\nu dud}^{\tL\tR(\tL\tL),y}$ through diagrams that involve weak neutral-current four-fermion vertices from both neutrino-neutron and neutrino-neutrino interactions. 

In the $\Delta(B+L)=0$ sector, 
the leading-order diagrams for the decay $n \to \ell^{-} \ell^{+} \nu$ closely resemble those for $n \to \ell^{-} \ell^{+} \bar\nu$, with the difference being the BNV operators involved. 
Specifically, the photon-mediated diagrams involve the operators $\calO_{\bar\nu dud}^{\tR\tL(\tR\tR),y}$, while the $K^0$- and $K^\pm$-mediated diagrams are induced by $\calO_{\bar\nu uds}^{\tR\tL,y},\calO_{\bar\nu dsu}^{\tR\tL(\tR\tR),y},\calO_{\bar\nu sud}^{\tR\tL(\tR\tR),y}$ and $\calO_{\bar\ell dds}^{{\tt XY},y}$, respectively. For the decay $p \to \ell^{+} \nu \nu$, the dominant contributions arise from the $\pi^\pm$- and $K^\pm$-mediated diagrams induced by the same set of operators as in $n \to \ell^{-} \ell^{+} \nu$
excluding $\calO_{\bar\ell dds}^{{\tt XY},y}$. 
Lastly, the processes in the class $n \to \nu \bar{\nu} \bar{\nu}$ are uniquely generated by the operators $\calO_{\bar\nu dud}^{\tR\tL(\tR\tR),y}$. 

%%%%%%%%%%%%%%%%%%%%%%%%%
\begin{table*}[t]
\centering
\setlength{\extrarowheight}{2pt}
\resizebox{0.95\linewidth}{!}{
\renewcommand{\arraystretch}{1.1}
\begin{tabular}{|c|l|c|c|c|}
\hline
\multicolumn{5}{|c|}{ $\Delta(B-L)=0$} 
\\\hline
\multirow{2}{*}{Mode}
& \multicolumn{1}{c|}{\multirow{2}{*}{Decay width expression $\Gamma\,[10^{-9}\,\GeV^5]$}} 
& \multicolumn{3}{c|}{Bounds on $\Gamma^{-1}\,[\rm yr]$}
\\\cline{3-5}
& &{\bf This work} & Experiment &~Ref.~\cite{Girmohanta:2019xya}~
\\\hline
$ p\to e^{-} e^{+} e^{+}$ 
& $\makecell[l]{~~\,2.35\left(\left|C^{\tL,e}_{\ell uud}\right|^2 
+ \left| C^{\tR,e}_{\ell uud}\right|^2\right)
-0.0045 \Re\left(C^{\tL,e}_{\ell uud}C^{\tR,e*}_{\ell uud}\right)
\\+(10^{9}{\rm GeV}^{-5})\,\Gamma_{p\to e^+ K^0}\left(C_{\ell usu}^{{\tt XY},e}\right)\,
{\cal B}_{K^0\to e^- e^+} +\cdots
}$
& $\makecell{2.3\cdot 10^{39}\,(\dagger)\\
5.7\cdot 10^{42}\,(\star)}$
& $3.4\cdot 10^{34}$~\cite{Super-Kamiokande:2020tor}
& \multirow{8}{*}{$10^{37}$}
\\\cline{1-4}
$ p\to e^{-} e^{+} \mu^{+}$ 
& $\makecell[l]{~~\,2.28\left(\left| C^{\tL,\mu}_{\ell uud}\right|^2
+ \left| C^{\tR,\mu}_{\ell uud}\right|^2\right)
-0.96 \Re\left(C^{\tL,\mu}_{\ell uud}C^{\tR,\mu*}_{\ell uud}\right)\\
+(10^{9}{\rm GeV}^{-5})\,\Gamma_{p\to \mu^+ K^0}\left(C_{\ell usu}^{{\tt XY},\mu}\right)\,
{\cal B}_{K^0\to e^- e^+}+\cdots}$
& $\makecell{1.6\cdot 10^{39}\,(\dagger)\\
9.2\cdot  10^{42}\,(\star)} $
&$2.3\cdot 10^{34}$~\cite{Super-Kamiokande:2020tor}
&
\\\cline{1-4}
$ p\to \mu^{-} \mu^{+} e^{+}$ 
& $\makecell[l]{~~\,0.28\left( \left| C^{\tL,e}_{\ell uud}\right|^2
+ \left| C^{\tR,e}_{\ell uud}\right|^2\right)
-0.00035 \Re\left(C^{\tL,e}_{\ell uud}C^{\tR,e*}_{\ell uud}\right)\\
+(10^{9}{\rm GeV}^{-5})\,\Gamma_{p\to e^+ K^0}\left(C_{\ell usu}^{{\tt XY},e}\right)\,
{\cal B}_{K^0\to \mu^- \mu^+}+\cdots}$
& $\makecell{1.9\cdot  10^{40}\,(\dagger)\\
7.5\cdot  10^{39}\,(\star) } $
&$9.2\cdot  10^{33}$~\cite{Super-Kamiokande:2020tor}
&  
\\\cline{1-4}
$ p\to \mu^{-} \mu^{+} \mu^{+}$ 
& $\makecell[l]{~~\,0.31\left( \left|C^{\tL,\mu}_{\ell uud}\right|^2 
+ \left|C^{\tR,\mu}_{\ell uud}\right|^2\right)
-0.052 \Re\left(C^{\tL,\mu}_{\ell uud}C^{\tR,\mu*}_{\ell uud}\right)\\
+(10^{9}{\rm GeV}^{-5})\,\Gamma_{p\to \mu^+ K^0}\left(C_{\ell usu}^{{\tt XY},\mu}\right)\,
{\cal B}_{K^0\to \mu^- \mu^+}+\cdots}$
& $\makecell{1.1\cdot 10^{40}\,(\dagger)\\
1.2\cdot 10^{40}\,(\star) } $
&$1.0\cdot 10^{34}$~\cite{Super-Kamiokande:2020tor}
&
\\\clineB{1-5}{2}
$\makecell{n\to e^- e^+ \bar{\nu}_{y}\\
\{y=(e),\mu,\tau\}}$
& $\makecell[l]{~~\,2.62\left|C_{\nu dud}^{\tL,y}\right|^2 \\
+\delta_{y,e}(10^{9}{\rm GeV}^{-5})\,\Gamma_{n\to e^+ \pi^-}\left(C_{\ell uud}^{{\tt XY},e}\right)\,
{\cal B}_{\pi^-\to e^-\bar\nu_e}\\
+(10^{9}{\rm GeV}^{-5})\,\Gamma_{n\to \bar\nu K^0}\left(C_{\nu uds}^{{\tt LR},y},C_{\nu dsu}^{{\tt LX},y},C_{\nu sud}^{{\tt LX},y}\right)\,
{\cal B}_{K^0\to e^- e^+}
+ \cdots }$
&$\makecell{9.3\cdot 10^{37}\,(\dagger) \\
(4.3)\cdot 10^{37}\,(\smallstar)\\
1.2\cdot 10^{46}\,(\star) }$
&$2.57\cdot 10^{32}$~\cite{McGrew:1999nd}
& \multirow{4}{*}{$10^{36}$}
\\\cline{1-4}
$\makecell{n\to \mu^- \mu^+ \bar{\nu}_{y}\\\{y=e,(\mu),\tau\}}$
& $ \makecell[l]{~~\,0.32\left|C_{\nu dud}^{\tL,y}\right|^2 \\
+\delta_{y,\mu}(10^{9}{\rm GeV}^{-5})\,\Gamma_{n\to \mu^+ \pi^-}\left(C_{\ell uud}^{{\tt XY},\mu}\right)\,
{\cal B}_{\pi^-\to \mu^-\bar\nu_\mu}\\
+(10^{9}{\rm GeV}^{-5})\,\Gamma_{n\to \bar\nu K^0}\left(C_{\nu uds}^{{\tt LR},y},C_{\nu dsu}^{{\tt LX},y},C_{\nu sud}^{{\tt LX},y}\right)\,
{\cal B}_{K^0\to \mu^- \mu^+}
+\cdots }$
&$\makecell{7.6\cdot 10^{38}\,(\dagger) \\
(3.5)\cdot 10^{33}\,(\filledstar)\\
5.1\cdot 10^{43}\,(\star) } $
&$7.9\cdot 10^{31}$~\cite{McGrew:1999nd}
& 
\\\hline
$ n\to e^{-} \mu^{+} \bar{\nu}_{e}$
&$\makecell[l]{~~\,271.7\times10^{-11}\left|C_{\nu dud}^{\tL,\mu}\right|^2 \\
+(10^{9}{\rm GeV}^{-5})\,\Gamma_{n\to \mu^+ \pi^-}\left(C_{\ell uud}^{{\tt XY},\mu}\right)\,
{\cal B}_{\pi^-\to e^-\bar\nu_e}
+ \cdots}$
&$\makecell{9.0\cdot 10^{46}\,(\diamond) \\
2.8\cdot 10^{37}\,(\smallstar)}$
&$8.3\cdot 10^{31}$~\cite{McGrew:1999nd}
&$10^{44}$
\\\cline{1-5}
$ n\to \mu^{-} e^{+} \bar{\nu}_{\mu} $
&$\makecell[l]{~~\,271.7\times10^{-11}\left|C_{\nu dud}^{\tL,e}\right|^2 \\
+ (10^{9}{\rm GeV}^{-5})\,\Gamma_{n\to e^+ \pi^-}\left(C_{\ell uud}^{{\tt XY},e}\right)\,
{\cal B}_{\pi^-\to \mu^-\bar\nu_\mu}
+\cdots}$
&$\makecell{9.0\cdot 10^{46}\,(\diamond) \\
5.3\cdot 10^{33}\,(\filledstar)}$ 
& $0.6\cdot 10^{30}$~\cite{Learned:1979gp}
& ---
\\\clineB{1-5}{2}
$\makecell{p\to e^{+} \nu_y \bar\nu_y\\
\{y=\mu,\tau\}}$
&~~\,$10^{-11} \Big[98.04\left(\left|C^{\tL,e}_{\ell uud}\right|^2 
+ \left| C^{\tR,e}_{\ell uud}\right|^2\right)
-0.43\Re\left(C^{\tL,e}_{\ell uud}C^{\tR,e*}_{\ell uud}\right)\Big]$
& $5.5\cdot 10^{48}\,(\diamond) $  
& \multirow{6}{*}{$1.7\cdot 10^{32}$~\cite{Super-Kamiokande:2014pqx}}
& ---
\\\hhline{|---~-|}
$\makecell{p\to e^{+} \nu_e \bar\nu_y\\
\{y=e,(\mu),[\tau]\}}$
&$\makecell[l]{~~\,
10^{-11}\Big[ 738(306)[44.6]\left|C^{\tL,y}_{\ell uud}\right|^2\\  
+ 98(3.9)[160]\left| C^{\tR,y}_{\ell uud}\right|^2 
-1.5(69)[169]\Re\left(C^{\tL,y}_{\ell uud}C^{\tR,y*}_{\ell uud}\right)\Big] 
\\
+ (10^{9}{\rm GeV}^{-5})\,\Gamma_{p\to \bar\nu_y \pi^+}\left(C_{\nu dud}^{\tL{\tt X},y}\right)\,{\cal B}_{\pi^+\to e^+\nu_e} \\
+ (10^{9}{\rm GeV}^{-5}) \Gamma_{p\to \bar\nu_y K^+}\left(C_{\nu uds}^{\tL\tR,y},C_{\nu dsu}^{\tL{\tt X},y},C_{\nu sud}^{\tL{\tt X},y}\right)\,{\cal B}_{K^+\to e^+\nu_e}+
\cdots}$  
& $\makecell{ 7.3(12)[-]\cdot 10^{47}\,(\diamond) 
\vspace{0.4em}\\
5.5(91)[-]\cdot 10^{48}\,(\diamond) 
\vspace{0.4em}\\
3.2\cdot 10^{36}\,(\smallstar)
\vspace{0.2em}\\
3.7\cdot 10^{38}\,(\smallstar) }$
&
&$10^{45}$
\\\hhline{|-----|}
$\makecell{p\to \mu^{+} \nu_y \bar\nu_y\\
\{y=e,\tau\}}$
&~~$10^{-11} \Big[96.34\left(\left|C^{\tL,\mu}_{\ell uud}\right|^2 
+ \left| C^{\tR,\mu}_{\ell uud}\right|^2\right)
-76.66\Re\left(C^{\tL,\mu}_{\ell uud}C^{\tR,\mu*}_{\ell uud}\right)\Big]$
& $3.7\cdot 10^{48}\,(\diamond) $  
& \multirow{6}{*}{$2.2\cdot 10^{32}$~\cite{Super-Kamiokande:2014pqx}}
& ---
\\\hhline{|---~-|}
$\makecell{p\to \mu^{+} \nu_\mu \bar\nu_y\\
\{y=e,(\mu),[\tau]\} }$
&$\makecell[l]{
~~10^{-11} \Big[270.5(699)[40.4]\left|C^{\tL,y}_{\ell uud}\right|^2\\ 
+ 8\cdot 10^{-5}(111)[145]\left| C^{\tR,y}_{\ell uud}\right|^2
-0.3(273)[153]\Re\left(C^{\tL,y}_{\ell uud}C^{\tR,y*}_{\ell uud}\right)\Big]
\\
+(10^{9}{\rm GeV}^{-5})\,\Gamma_{p\to \bar\nu_y \pi^+}\left(C_{\nu dud}^{\tL{\tt X},y}\right)\,{\cal B}_{\pi^+\to \mu^+\nu_\mu} \\
+ (10^{9}{\rm GeV}^{-5}) \Gamma_{p\to \bar\nu_y K^+}\left(C_{\nu uds}^{\tL\tR,y},C_{\nu dsu}^{\tL{\tt X},y},C_{\nu sud}^{\tL{\tt X},y}\right)\,{\cal B}_{K^+\to \mu^+\nu_\mu}+\cdots}$~~  
& $\makecell{ 20(5.1)[-]\cdot 10^{47}\,(\diamond)  
\vspace{0.4em}\\
~67(32)[-]\cdot 10^{53(47)}\,(\diamond) 
\vspace{0.4em}\\
3.9\cdot 10^{32}\,(\filledstar)
\vspace{0.2em}\\
9.3\cdot 10^{33}\,(\filledstar) }$
&
&$10^{45}$
\\\clineB{1-5}{2}
$\makecell{n\to\nu_{x}\bar{\nu}_{x}\bar{\nu}_{y}\\~\{x,y=e,\mu,\tau\}~~}$
&~~$92.8(185.6)\cdot 10^{-11}\left| C_{\nu dud}^{\tL,y}\right|^2\text{~for~}x\neq y(x=y)$
& $2.6(1.3)\cdot 10^{47}\,(\diamond)$
&~$0.9\cdot 10^{30}$~\cite{SNO:2022trz}~
&$10^{44}$
\\
\hline
\end{tabular} }
\caption{Summary of the nucleon triple-lepton decay widths (in the second column) induced by dim-6 BNV operators in the LEFT, together with the newly derived lower bounds on their partial lifetimes (in the third column). The dependence on WCs of the two-body decay widths is explicitly indicated in brackets. The ellipses denote interference terms between operators with different field configurations. 
The values marked with a dagger ($\dagger$), a diamond ($\diamond$), and a star ($\star/\smallstar/\filledstar$) represent the bounds resulting from the photon-mediated, four-fermion-involved, and meson-mediated contributions, respectively. 
For comparison, the fourth column presents the experimental bounds, and the last column is the estimates based on phase space difference and counting of SM couplings between two- and three-body decays in~\cite{Girmohanta:2019xya}. }
\label{tab:ResBmL=0}
\end{table*}

%%%%%%%%%%%%%%%%%%%%%%%%%
\begin{table*}[t]
\centering
\setlength{\extrarowheight}{2pt}
\resizebox{0.80\linewidth}{!}{
\renewcommand{\arraystretch}{1.7}
\begin{tabular}{|c|l|c|c|}
\hline
\multicolumn{4}{|c|}{$\Delta(B+L)=0$} 
\\\hline
\multirow{2}{*}{Mode}
& \multicolumn{1}{c|}{\multirow{2}{*}{Decay width expression $\Gamma\,[10^{-9}\,\GeV^5]$}} 
& \multicolumn{2}{c|}{Bounds on $\Gamma^{-1}\,[\rm yr]$}
\\\cline{3-4}
& & {\bf This work} & Experiment 
\\\cline{1-4}
$\makecell{ n\to e^{-}  e^{+} \nu_{y}\\
\{y=(e),\mu,\tau\}}$
&$\makecell[l]{~~\,2.62\left|C_{\bar{\nu} dud}^{\tR,y}\right|^2\\
+ \delta_{y,e} (10^{9}{\rm GeV}^{-5})\,\Gamma_{n\to e^- K^+}\left(C_{\bar\ell dds}^{{\tt XY},e}\right)\,{\cal B}_{K^+\to e^+\nu_e}\\
+(10^{9}{\rm GeV}^{-5})\,\Gamma_{n\to \nu K^0}\left(C_{\nu uds}^{{\tt RL},y},C_{\nu dsu}^{{\tt RX},y},C_{\nu sud}^{{\tt RX},y}\right)\,
{\cal B}_{K^0\to e^- e^+}+ \cdots }$
& $\makecell{9.3\cdot 10^{37}\,(\dagger) \\
(2.0)\cdot 10^{36}\,(\smallstar) \\
1.2\cdot 10^{46}\,(\star) } $
& $2.57\cdot 10^{32}$~\cite{McGrew:1999nd}
\\\cline{1-4}
$\makecell{n\to \mu^{-}  \mu^{+} \nu_{y}\\\{y=e,(\mu),\tau\}}$
&$\makecell[l]{~~\,0.32\left|C_{\bar{\nu} dud}^{\tR,y}\right|^2\\
+ \delta_{y,\mu}(10^{9}{\rm GeV}^{-5})\,\Gamma_{n\to \mu^- K^+}\left(C_{\bar\ell dds}^{{\tt XY},\mu}\right)\,{\cal B}_{K^+\to \mu^+\nu_\mu}\\
+(10^{9}{\rm GeV}^{-5})\,\Gamma_{n\to \nu K^0}\left(C_{\nu uds}^{{\tt RL},y},C_{\nu dsu}^{{\tt RX},y},C_{\nu sud}^{{\tt RX},y}\right)\,
{\cal B}_{K^0\to \mu^- \mu^+}
+ \cdots}$
& $\makecell{7.6\cdot 10^{38}\,(\dagger) \\
(9.0)\cdot 10^{31}\,(\filledstar)\\
5.1\cdot 10^{43}\,(\star) } $
& $7.9\cdot 10^{31}$~\cite{McGrew:1999nd}
\\\cline{1-4}
$ n\to e^{-} \mu^{+} \nu_{\mu}$
&$\makecell[l]{~~\,271.7\times10^{-11}\left|C_{\bar\nu dud}^{\tR,e}\right|^2\\
+ (10^{9}{\rm GeV}^{-5})\,\Gamma_{n\to e^- K^+}\left(C_{\bar\ell dds}^{{\tt XY},e}\right)\,{\cal B}_{K^+\to \mu^+\nu_\mu}
+ \cdots}$
& $\makecell{9.0\cdot 10^{46}\,(\diamond)  \\
 5.0\cdot 10^{31}\,(\filledstar) }$
& $8.3\cdot 10^{31}$~\cite{McGrew:1999nd}
\\\cline{1-4}
$ n\to \mu^{-} e^{+} \nu_e $
&$\makecell[l]{~~\,271.7\times10^{-11}\left|C_{\bar\nu dud}^{\tR,\mu}\right|^2\\
+ (10^{9}{\rm GeV}^{-5})\,\Gamma_{n\to \mu^- K^+}\left(C_{\bar\ell dds}^{{\tt XY},\mu}\right)\,{\cal B}_{K^+\to e^+\nu_e}
+ \cdots}$
& $\makecell{9.0\cdot 10^{46}\,(\diamond)  \\
3.6\cdot 10^{36}\,(\smallstar) }$
& $0.6\cdot 10^{30}$~\cite{Learned:1979gp}
\\\clineB{1-4}{2}
$\makecell{p\to e^{+} \nu_e \nu_y\\
\{y=e,\mu,\tau\}}$  
&$\makecell[l]{~~\,(10^{9}{\rm GeV}^{-5})\,\Gamma_{p\to \nu_y \pi^+} \left(C_{\bar\nu dud}^{\tR{\tt X},y}\right)\,{\cal B}_{\pi^+\to e^+\nu_e} \\
+ (10^{9}{\rm GeV}^{-5}) \Gamma_{p\to \nu_y K^+}\left(C_{\bar\nu uds}^{\tR\tL,y},C_{\bar\nu dsu}^{\tR{\tt X},y},C_{\bar\nu sud}^{\tR{\tt X},y}\right)\,{\cal B}_{K^+\to e^+\nu_e}+\cdots}$ 
& $\makecell{3.2\cdot 10^{36}\,(\smallstar) \\
3.7\cdot 10^{38}\,(\smallstar) }$
& \multirow{1}{*}{$1.7\cdot 10^{32}$~\cite{Super-Kamiokande:2014pqx}}
\\\cline{1-4}
$\makecell{p\to \mu^{+}  \nu_\mu \nu_y\\
\{y=e,\mu,\tau\}}$  
&$\makecell[l]{~~\,(10^{9}{\rm GeV}^{-5})\,\Gamma_{p\to \nu_y \pi^+}\left(C_{\bar\nu dud}^{\tR{\tt X},y}\right)\,{\cal B}_{\pi^+\to \mu^+\nu_\mu} \\
+ (10^{9}{\rm GeV}^{-5}) \Gamma_{p\to \nu_y K^+}\left(C_{\bar\nu uds}^{\tR\tL,y},C_{\bar\nu dsu}^{\tR{\tt X},y},C_{\bar\nu sud}^{\tR{\tt X},y}\right)\,{\cal B}_{K^+\to \mu^+\nu_\mu}+\cdots}~~$  
& $\makecell{3.9\cdot 10^{32}\,(\filledstar) \\
9.3\cdot 10^{33}\,(\filledstar)}$
& \multirow{1}{*}{$2.2\cdot 10^{32}$~\cite{Super-Kamiokande:2014pqx}}
\\\clineB{1-4}{2}
$\makecell{n\to\bar{\nu}_{x}\nu_{x}\nu_{y}\\
~~\{x,y=e,\mu,\tau\}~~} $
&~~$334.4 (668.8)\cdot 10^{-11} \left|C_{\bar{\nu} dud}^{\tR,y}\right|^2\text{~for~}x\neq y(x=y)$
&~~$7.3(3.7)\cdot 10^{46}\,(\diamond) $~~
&~~$0.9\cdot 10^{30}$~\cite{SNO:2022trz}~~
\\
\hline
\end{tabular} }
\caption{Same as \cref{tab:ResBmL=0} but for the $\Delta(B+L)=0$ sector.}
\label{tab:ResBpL=0}
\end{table*}

%%%%%%%%%%%%%%%%%%%%%%%%%%%%%%%%%%%%%%
\section{Results}
\label{sec:results}
%%%%%%%%%%%%%%%%%%%%%%%%%%%%%%%%%%%%%%

Having identified the leading-order diagrams and the relevant BNV operators in the previous section, we now proceed to calculate their decay widths and derive new  bounds on their occurrence.
For a generic nucleon triple-lepton decay mode, ${\tt N}(p)\to l_1(p_1) l_2(p_2) l_3(p_3)$, 
the decay amplitude associated with the retained diagrams in \cref{tab:LO-diagrams} can be directly written down using the vertices provided in \cref{app:BNV-ChPT,app:BNC-ChPT}. The corresponding decay width is then calculated as
\begin{align}
\label{eq:decaywidth}
\Gamma=\frac{1}{S!}\frac{1}{256\pi^{3}}\frac{1}{m_{\N}^3}\int{ds} \int_{t_-}^{t_+}{dt} ~\overline{|\calM|^2},
\end{align}
where $\overline{|\calM|^2}$ denotes the spin-averaged and -summed matrix element squared, and $S(=1,2)$ is a symmetry factor accounting for identical particles in the final state.
The kinematically allowed integration regions are specified by
\begin{align}
&(m_{1}+m_{2})^2\leq s\leq (m_{\N}-m_{3})^2,\nonumber
\\
t_{\pm} & = (E_{2}^{*}+E_{3}^{*})^2-\Big(\sqrt{E_{2}^{*2}-m_{2}^2}\mp \sqrt{E_{3}^{*2}-m_{3}^2}\Big)^2,\nonumber
\\
E_{2}^{*} &=\frac{s-m_{1}^2+m_{2}^2}{2\sqrt{s}},\, E_{3}^{*}=\frac{m_{\N}^2-s-m_{3}^2}{2\sqrt{s}}.
\end{align}
Here, $s\equiv (p_1+p_2)^2$ and $t\equiv (p_2+p_3)^2 $, where $p_i$ represents the four-momentum of the $i$-th final-state lepton and $m_i$ denotes its mass. 
In calculating each decay width, interference terms between operators differing in lepton and quark flavor configurations are neglected. However, interference terms between operators sharing the same flavor combinations but differing in chiral structures are retained.

In practice, the purely meson-mediated contributions experience resonance enhancement and can be factorized, using the narrow-width approximation method, into the product of the relevant two-body nucleon decay width and the corresponding meson decay branching ratio, i.e., 
$\Gamma_{{\tt N} \to l_1 l_2 l_3}(C_i) = \Gamma_{{\tt N} \to M l_1}(C_i) {\cal B}_{M \to l_2 l_3}$. 
Here, $C_i$ denote the WCs of the corresponding dim-6 BNV operators that induce both the two- and three-body decays. 
As we will discuss below, these resonance effects are crucial for correctly deriving bounds on these nucleon triple-lepton decay modes.

We formulate the decay widths in terms of the WCs of the relevant dim-6 BNV operators. To perform the numerical integration over the phase space, we use the central values of SM parameters from the latest PDG~\cite{ParticleDataGroup:2024cfk}, and the relevant hadronic LECs appearing in \cref{Eq.BNV-ChPT,eq:Bchpt} are taken from~\cite{Yoo:2021gql,Bali:2022qja}.
The branching ratios for the two-body leptonic decays of the mesons associated with the meson-mediated diagrams are summarized in \cref{tab:data}.
Our key results for the final decay widths are summarized in the second column of \cref{tab:ResBmL=0} and \cref{tab:ResBpL=0}, corresponding to the $\Delta(B-L)=0$ and $\Delta(B+L)=0$ sectors, respectively. 
Contributions from photon-mediated diagrams and those involving four-fermion vertices are directly parametrized by the relevant WCs, whereas meson-mediated contributions are expressed as the product of the two-body nucleon decay widths and the corresponding meson leptonic decay branching ratios.
For the former, the following combinations of WCs are adopted
\begin{align}
C_{\ell uud}^{\tL,x}\equiv C_{\ell uud}^{\tL\tR,x} +\kappa C_{\ell uud}^{\tL\tL,x},\,
C_{\nu dud}^{\tL,x} \equiv C_{\nu dud}^{\tL\tR,x} +\kappa C_{\nu dud}^{\tL\tL,x},
\end{align}
together with their chirality partners obtained by $\tL\leftrightarrow \tR$ (and $\nu\leftrightarrow \bar\nu$). Here, $\kappa\equiv c_2/c_1\approx-1.0096$.
For the latter, the analytical expressions of the relevant two-body nucleon decays can be found in the previous paper~\cite{Liao:2025sqt}. 
Using the narrow-width approximation, we estimate that the $\rho^0$- and $\omega$-mediated contributions to the decay widths of nucleon modes with an $e^-e^+$ pair are approximately one order of magnitude smaller than the photon-mediated contributions. For the decay modes with a $\mu^-\mu^+$ pair, the former contributions are at most comparable to the latter.

Using the current experimental bounds on $\Gamma^{-1}$s of the two-body nucleon decays and the meson leptonic decay branching ratios listed in~\cref{tab:data}, together with the constraints on the WCs of the relevant dim-6 operators presented in Table\,S4 of the Supplemental Material of Ref.\,\cite{Liao:2025sqt}, 
we derive new lower bounds on these $\Delta F_L=1$  nucleon decays into three leptons. 
Our final results are shown in the third column of \cref{tab:ResBmL=0,tab:ResBpL=0}. 
The dependence of the derived bounds on the WCs of dim-6 operators is explicitly indicated in the tables.
Note that for resonance contributions involving $K^0$, the meson decay branching ratios are given in terms of their mass eigenstates, $K_L$ and $K_S$. Therefore, we use the experimental limits on nucleon decays involving $K_{L,S}$ to obtain conservative estimates. 
Although we have neglected the contributions mediated by vector mesons as well as hadronic uncertainties, we expect these effects to modify the derived bounds by at most an $\calO(1)$ factor.

It can be seen that our derived bounds on $\Gamma^{-1}$ from photon-mediated contributions are on the order of $\calO(10^{38-40}\,\rm yr)$  (labeled by $\dagger$), while those involving weak four-fermion vertices are typically around $\calO(10^{46-49}\,\rm yr)$ (labeled by $\diamond$).
However, the bounds on decay modes involving muon-flavor lepton pairs $\mu^+\nu_\mu
$ and $\mu^-\bar\nu_\mu$ from the $\pi^\pm$- and $K^\pm$-mediated diagrams are approximately $\calO(10^{32-34}\,\rm yr)$ (labeled by $\filledstar$), significantly weaker than others due to their large branching ratios. 
The bounds on decay modes involving electron-flavor lepton pairs $e^+\nu_e$ and $e^-\bar\nu_e$ from the same $\pi^\pm$- and $K^\pm$-mediated diagrams are around $\calO(10^{36-38}\,\rm yr)$ (labeled by $\smallstar$), as a result of their much smaller branching ratios. Finally, the bounds from $K^0$-mediated diagrams lie in the range of $\calO(10^{40-46}\,\rm yr)$ (labeled by $\star$), owing to their extremely small branching ratios.  

Finally, we make a few remarks on our results. 
First, Ref.~\cite{Girmohanta:2019xya} made an estimate of triple-lepton decay widths from two-body nucleon decays by attaching electroweak couplings and considering differences in two- and three-body phase spaces while treating decay amplitudes essentially as a constant. Its numbers are shown in the last column of \cref{tab:ResBmL=0}. In contrast, our results exhibit strong dependence on specific operators under consideration, with values varying by several orders of magnitude. In particular, the resonance effects in the meson-mediated contributions were totally neglected in \cite{Girmohanta:2019xya}.
Second, from the last three columns of \cref{tab:ResBmL=0}, we find that our bounds on the four decay modes of the proton into three charged leptons are at least 5 orders of magnitude more stringent than the current experimental limits. Moreover, these bounds improve upon previous estimates by 2 to 3 orders of magnitude. 
Third, for the remaining nucleon decay modes that do not involve a $\mu^+\nu_\mu$ or $\mu^-\bar\nu_\mu$ pair, our results surpass the existing experimental bounds by 4 to 17 orders of magnitude, depending on the processes and  dim-6 operators under consideration. 
For decay modes involving a $\mu^+\nu_\mu$ or $\mu^-\bar\nu_\mu$ pair, our derived bounds are comparable to their experimental limits.  
Finally, in realistic ultraviolet-complete models, the WCs of operators with different flavor structures may be correlated. In such scenarios, the conservative bounds derived from our operator-by-operator analysis could be modified to some extent, depending on specific patterns of these correlations.

%%%%%%%%%%%%%%%%%%%%%%%%
\section{Summary}
\label{sec:summary}
%%%%%%%%%%%%%%%%%%%%%%%%

We have systematically investigated the noncontact contributions to the $\Delta F_{L}=1$ nucleon triple-lepton decays induced by dim-6 BNV operators through SM mediators. Employing chiral perturbation theory, we collected at leading order all relevant hadronic-level interactions responsible for these decay modes. For each decay mode, we analyzed the leading-order Feynman diagrams that are induced by various dim-6 BNV operators involving light $u,d,s$ quarks and calculated its decay width. 
By combining experimental limits on nucleon two-body decays with existing bounds on these operators from the literature, we established new stringent lower limits on the partial lifetimes of the triple-lepton decay modes. 
The formulas for the decay widths and our new bounds are summarized in \cref{tab:ResBmL=0,tab:ResBpL=0}.

Our results indicate that the meson-mediated diagrams involving a $\mu^+\nu_\mu$ or $\mu^-\bar\nu_\mu$ pair can be enhanced by resonance effects, leading to lifetime bounds from certain operators that are comparable to the current experimental limits. In contrast, the photon-mediated diagrams, the diagrams involving weak four-fermion interactions, and other meson-mediated contributions yield bounds that surpass existing experimental constraints by a factor ranging from $\calO(10^4)$ to $\calO(10^{17})$, depending on the specific mode and operators involved. 
These findings imply that noncontact contributions to nucleon triple-lepton decays induced by dim-6 operators are generally small, except for a few modes where resonance effects play a significant role. 
In light of this, we will consider the contact contributions from dim-9 operators in a forthcoming paper~\cite{Liao:2025wxk}. 

%%%%%%%%%%%%%%%%%%%%%%%%
\acknowledgments
%%%%%%%%%%%%%%%%%%%%%%%%
This work was supported 
by the Grants 
No.\,NSFC-12035008
and No.\,NSFC-12305110.

\onecolumngrid
\appendix

\newlength{\fwidth}
\setlength{\fwidth}{0.3\textwidth}

%%%%%%%%%%%%%%%%%%%%%%%%%%%%%%%%%%%%%%
\section{Summary of SM weak four-fermion interactions}
\label{app:SMweak}
%%%%%%%%%%%%%%%%%%%%%%%%%%%%%%%%%%%%%%

Many triple-lepton decay modes of the nucleon involve the SM leptonic and semi-leptonic weak interactions.  
In the semi-leptonic interactions, only the $u,d,s$ quarks are relevant here. These interactions are summarized in \cref{tab:SMweak}.
Note that the neutral current ($\tt NC$) and charged current ($\tt CC$) contributions to the left-handed charged lepton and neutrino interactions are explicitly separated.

%%%%%%%%%%%%%%%
\begin{table}[h]
\centering
\resizebox{0.6\linewidth}{!}{
\renewcommand{\arraystretch}{1.1}
\begin{tabular}{|c|c|c|l|}
\hline
~~~~& \multicolumn{1}{c|}{~Notation~~} &\multicolumn{1}{c|}{Operator}&  \multicolumn{1}{c|}{Matching coefficient} 
\\
\hline%%
~\multirow{14}{*}{\rotatebox[origin=c]{90}{Semileptonic} }
& $\mathcal{O}_{u\nu }^{\tL\tL}$ 
& $(\bar u_{\tL}\gamma_\mu u_{\tL})(\bar\nu_{\tL x}\gamma^\mu\nu_{\tL x})$ 
& $-4\sqrt{2}(\frac{1}{2}-\frac{2}{3}s_W^2)(\frac{1}{2})G_F$
\\
\cline{2-4}
& $\mathcal{O}_{ u\ell}^{\tL\tL}$ 
& $(\bar u_{\tL}\gamma_\mu u_{\tL})(\bar\ell_{\tL x}\gamma^\mu\ell_{\tL x})$ 
&$-4\sqrt{2}(\frac{1}{2}-\frac{2}{3}s_W^2) (-\frac{1}{2} + s_W^2)G_F$
\\
\cline{2-4}
& $\mathcal{O}_{u\ell}^{\tL\tR}$  
&$(\bar u_{\tL}\gamma^\mu u_{\tL})(\bar\ell_{\tR x}\gamma_\mu\ell_{\tR x})$
&$-4\sqrt{2} (\frac{1}{2}-\frac{2}{3}s^2_W)(s^2_W) G_F$ 
\\
\cline{2-4}%
& $\mathcal{O}_{u\nu }^{\tR\tL}$  
&$(\bar u_{\tR}\gamma_\mu u_{\tR})(\bar\nu_{\tL x}\gamma^\mu\nu_{\tL x})$
&$-4\sqrt{2}(-\frac{2}{3}s_W^2)(\frac{1}{2}) G_F$
\\
\cline{2-4}
& $\mathcal{O}_{u\ell}^{\tR\tL}$ 
&$(\bar u_{\tR}\gamma_\mu u_{\tR})(\bar\ell_{\tL x}\gamma^\mu\ell_{\tL x})$ 
&$-4\sqrt{2}(-\frac{2}{3}s_W^2)(-\frac{1}{2}+s_W^2)G_F$
\\
\cline{2-4}
& $\mathcal{O}_{u\ell}^{\tR\tR} $ 
&$(\bar u_{\tR}\gamma_\mu u_{\tR})(\bar\ell_{\tR x}\gamma^\mu\ell_{\tR x})$
&$-4\sqrt{2}(-\frac{2}{3}s_W^2)(s_W^2)G_F$
\\
\cline{2-4}%%
& $\mathcal{O}_{d\nu}^{\tL\tL} $ 
&$(\bar d_{\tL i}\gamma_\mu d_{\tL i})(\bar\nu_{\tL x}\gamma^\mu\nu_{\tL x})$   &$-4\sqrt{2}(-\frac{1}{2}+\frac{1}{3}s_W^2)(\frac{1}{2}) G_F$
\\
\cline{2-4}
& $\mathcal{O}_{ d\ell}^{\tL\tL}$  
&$(\bar d_{\tL i}\gamma_\mu d_{\tL i})(\bar\ell_{\tL x}\gamma^\mu\ell_{\tL x})$  
&$-4\sqrt{2}(-\frac{1}{2}+\frac{1}{3}s_W^2) (-\frac{1}{2} + s_W^2)G_F$
\\
\cline{2-4}
& $\mathcal{O}_{d\ell}^{\tL\tR} $ 
&$(\bar d_{\tL i}\gamma^\mu d_{\tL i})(\bar\ell_{\tR x}\gamma_\mu\ell_{\tR x})$
&
$-4\sqrt{2} (-\frac{1}{2}+\frac{1}{3}s_W^2)(s^2_W) G_F$
\\
\cline{2-4}%
& $\mathcal{O}_{d\nu}^{\tR\tL}$ 
&$(\bar d_{\tR i}\gamma_\mu d_{\tR i})(\bar\nu_{\tL x}\gamma^\mu\nu_{\tL x})$  
&$-4\sqrt{2}(\frac{1}{3}s_W^2)(\frac{1}{2}) G_F$
\\
\cline{2-4}
& $\mathcal{O}_{d\ell}^{\tR\tL} $ 
&$(\bar d_{\tR i}\gamma_\mu d_{\tR i})(\bar\ell_{\tL x}\gamma^\mu\ell_{\tL x})$
&$-4\sqrt{2}(\frac{1}{3}s_W^2)(-\frac{1}{2}+s_W^2)G_F$
\\
\cline{2-4}
& $\mathcal{O}_{d\ell}^{\tR\tR} $ 
&$(\bar d_{\tR i}\gamma_\mu d_{\tR i})(\bar\ell_{\tR x}\gamma^\mu\ell_{\tR x})$ 
&$-4\sqrt{2}(\frac{1}{3}s_W^2)(s_W^2)G_F$
\\
\cline{2-4}%%
& $\mathcal{O}_{ d u\nu \ell}^{\tL\tL,\tt CC}$ 
&$(\bar d_{\tL i}\gamma_\mu u_{\tL})(\bar\nu_{\tL x}\gamma^\mu\ell_{\tL x})$ 
&$-2\sqrt{2}G_FV_{ui}^{*} $
\\
\cline{2-4}
& $\mathcal{O}_{u d \ell \nu}^{\tL\tL,\tt CC} $ 
&$(\bar u_{\tL}\gamma_\mu d_{\tL i})(\bar\ell_{\tL x}\gamma^\mu\nu_{\tL x})$ 
&$-2\sqrt{2}G_FV_{ui}$
\\  
\hline
~\multirow{4}{*}{\rotatebox[origin=c]{90}{ Leptonic} }
& $\mathcal{O}_{\nu \nu }^{\tL\tL}$ 
&$(\bar\nu_{\tL x}\gamma^\mu\nu_{\tL x})(\bar\nu_{\tL y}\gamma_\mu\nu_{\tL y})$
&$-2\sqrt{2}(\frac{1}{2})(\frac{1}{2})G_F$
\\ 
\cline{2-4}
& $\mathcal{O}_{\nu \ell }^{\tL\tL,\tt NC}$ 
&$(\bar\nu_{\tL x}\gamma^\mu\nu_{\tL x})(\bar\ell_{\tL y}\gamma_\mu\ell_{\tL y})$ 
&$-4\sqrt{2} (\frac{1}{2})(-\frac{1}{2}+s_W^2)G_F$
\\ 
\cline{2-4}
& $\mathcal{O}_{\nu\ell\ell\nu }^{\tL\tL,\tt CC}$ 
&$(\bar\nu_{\tL x}\gamma^\mu\ell_{\tL x})(\bar\ell_{\tL y}\gamma_\mu\nu_{\tL y})$ &$-2\sqrt{2} G_F$
\\ 
\cline{2-4}
& $\mathcal{O}_{\nu \ell }^{\tL\tR}$ 
&$(\bar\nu_{\tL x}\gamma^\mu\nu_{\tL x})(\bar\ell_{\tR y}\gamma_\mu\ell_{\tR y})$ &$-4\sqrt{2}(\frac{1}{2})(s_W^2)G_F$
\\
\hline
\end{tabular} }
\caption{Semi-leptonic and leptonic four-fermion operators in the LEFT and their coefficients originating from the exchange of $W$ and $Z$ bosons in the SM, where $x,y=e,\mu,\tau$ and $i=d,s$. $G_F$ denotes the Fermi constant, and $V_{ij}$ are the CKM matrix elements.
For the four-neutrino operator $\calO_{\nu\nu}^{\tL\tL}$, both indices $x$ and $y$ run over the three neutrino flavors $e,\mu,\tau$. }
\label{tab:SMweak}
\end{table}

%%%%%%%%%%%%%%%%%%%%%%%%%%%%%%%%%%%%%%
\section{The relevant BNV interactions at hadronic level}
\label{app:BNV-ChPT}
%%%%%%%%%%%%%%%%%%%%%%%%%%%%%%%%%%%%%%

Denoting 
${\cal N}_{yzw}^{\tL\tL} \equiv  q_{\tL, y}^\alpha (\overline{ q_{\tL, z}^{\beta \C} } q_{\tL, w}^\gamma )\epsilon_{\alpha \beta \gamma}$
and ${\cal N}_{yzw}^{\tR\tL} \equiv q_{\tR, y}^\alpha (\overline{ q_{\tL, z}^{\beta \C} } q_{\tL,w}^\gamma)\epsilon_{\alpha \beta \gamma}$,
the matrices ${\cal N}_{{\bf 8}_\tL\otimes {\bf 1}_\tR}$ and 
${\cal N}_{\bar{\pmb{3}}_\tL \otimes \pmb{3}_\tR }$
appearing in \cref{eq:LEFTBNVLag} are the triple-quark components in the irreps ${\bf 8}_\tL\otimes {\bf 1}_\tR$ and $\bar{\pmb{3}}_\tL \otimes \pmb{3}_\tR$~\cite{Fan:2024gzc}:
\begin{align}
{\cal N}_{{\bf 8}_\tL\otimes {\bf 1}_\tR}
=
\begin{pmatrix}
{\cal N}^{\tL\tL}_{uds}  &  {\cal N}^{\tL\tL}_{usu}  & {\cal N}^{\tL\tL}_{uud}  
\\[1pt]
{\cal N}^{\tL\tL}_{dds}  & {\cal N}^{\tL\tL}_{dsu} & {\cal N}^{\tL\tL}_{dud}  
\\[1pt]
{\cal N}^{\tL\tL}_{sds} & {\cal N}^{\tL\tL}_{ssu} & {\cal N}^{\tL\tL}_{sud}
\end{pmatrix}, \quad 
{\cal N}_{\bar{\pmb{3}}_\tL \otimes \pmb{3}_\tR } = 
 \begin{pmatrix}
{\cal N}_{uds}^{\tR\tL}  & {\cal N}_{usu}^{\tR\tL} & {\cal N}_{uud}^{\tR\tL} 
\\[1pt]
{\cal N}_{dds}^{\tR\tL}  & {\cal N}_{dsu}^{\tR\tL} & {\cal N}_{dud}^{\tR\tL} 
\\[1pt] 
{\cal N}_{sds}^{\tR\tL}  & {\cal N}_{ssu}^{\tR\tL} & {\cal N}_{sud}^{\tR\tL}
\end{pmatrix}.
\label{eq:3qpart}
\end{align}
Their chirality partners are obtained by interchanging the left- and right-handed chiralities, $\tL\leftrightarrow\tR$.
The corresponding spurion field matrices are 
\begin{align}
{\cal P}_{\pmb{8}_\tL \otimes \pmb{1}_\tR} = &
\begin{pmatrix}
0  &  {\cellcolor{gray!15} C^{\tL\tL,x}_{\bar{\ell} dds} \overline{\ell_{\tR x}} }
& {\cellcolor{gray!15} C^{\tL\tL,x}_{\bar{\ell} sds} \overline{\ell_{\tR x}} } 
\\[4pt]
C^{\tL\tL,x}_{\ell usu} \overline{\ell_{\tL x}^{\C}} 
& C^{\tL\tL,x}_{\nu dsu} \overline{\nu_{\tL x}^{\C}} 
& C^{\tL\tL,x}_{\nu ssu} \overline{\nu_{\tL x}^{\C}} 
\\[4pt]
C^{\tL\tL,x}_{\ell uud} \overline{\ell_{\tL x}^{\C}} 
& C^{\tL\tL,x}_{\nu dud} \overline{\nu_{\tL x}^{\C}} 
& C^{\tL\tL,x}_{\nu sud} \overline{\nu_{\tL x}^{\C}} 
\end{pmatrix}, 
&
{\cal P}_{\pmb{1}_\tL \otimes \pmb{8}_\tR} = &
\begin{pmatrix}
0  &  {\cellcolor{gray!15} C^{\tR\tR,x}_{\bar{\ell} dds} \overline{\ell_{\tL x}} }
& {\cellcolor{gray!15} C^{\tR\tR,x}_{\bar{\ell} sds} \overline{\ell_{\tL x}} } 
\\[4pt]
C^{\tR\tR,x}_{\ell usu} \overline{\ell_{\tR x}^{\C}} 
& {\cellcolor{gray!15} C^{\tR\tR,x}_{\bar{\nu} dsu}  \overline{\nu_{\tL x}} }
& {\cellcolor{gray!15} C^{\tR\tR,x}_{\bar{\nu} ssu}  \overline{\nu_{\tL x}} } 
\\[4pt]
C^{\tR\tR,x}_{\ell uud} \overline{\ell_{\tR x}^{\C}} 
& {\cellcolor{gray!15} C^{\tR\tR,x}_{\bar{\nu} dud}  \overline{\nu_{\tL x}} }
& {\cellcolor{gray!15} C^{\tR\tR,x}_{\bar{\nu} sud} \overline{\nu_{\tL x}} }
\end{pmatrix},
\nonumber\\
{\cal P}_{\pmb{3}_\tL \otimes \bar{\pmb{3}}_\tR} = &
\begin{pmatrix}	
{\cellcolor{gray!15} C^{\tR\tL,x}_{\bar{\nu}uds}\overline{\nu_{\tL x}}  } 
& {\cellcolor{gray!15} C^{\tR\tL,x}_{\bar{\ell} dds} \overline{\ell_{\tL x}} }
& {\cellcolor{gray!15} C^{\tR\tL,x}_{\bar{\ell} sds} \overline{\ell_{\tL x}} }
\\[4pt]
C^{\tR\tL,x}_{\ell usu} \overline{\ell_{\tR x}^{\C}}
& {\cellcolor{gray!15} C^{\tR\tL,x}_{\bar{\nu} dsu}  \overline{\nu_{\tL x}} }
& {\cellcolor{gray!15} C^{\tR\tL,x}_{\bar{\nu} ssu}  \overline{\nu_{\tL x}} }
\\[4pt]
C^{\tR\tL,x}_{\ell uud} \overline{\ell_{\tR x}^{\C}} 
&{\cellcolor{gray!15} C^{\tR\tL,x}_{\bar{\nu} dud}  \overline{\nu_{\tL x}} }
&{\cellcolor{gray!15} C^{\tR\tL,x}_{\bar{\nu} sud} \overline{\nu_{\tL x}} }
\end{pmatrix}, 
&
{\cal P}_{\bar{\pmb{3}}_\tL \otimes \pmb{3}_\tR} = &
\begin{pmatrix}
C^{\tL\tR,x}_{\nu uds}\overline{\nu_{\tL x}^{\C}} &
{\cellcolor{gray!15} C^{\tL\tR,x}_{\bar{\ell} dds} \overline{\ell_{\tR x}} }
&{\cellcolor{gray!15} C^{\tL\tR,x}_{\bar{\ell}sds} \overline{\ell_{\tR x}} } 
\\[4pt]
C^{\tL\tR,x}_{\ell usu} \overline{\ell_{\tL x}^{\C}}
& C^{\tL\tR,x}_{\nu dsu}  \overline{\nu_{\tL x}^{\C}} 
&C^{\tL\tR,x}_{\nu ssu}  \overline{\nu_{\tL x}^{\C}} 
\\[4pt]
C^{\tL\tR,x}_{\ell uud} \overline{\ell_{\tL x}^{\C}} 
& C^{\tL\tR,x}_{\nu dud}  \overline{\nu_{\tL x}^{\C}} 
&C^{\tL\tR,x}_{\nu sud} \overline{\nu_{\tL x}^{\C}} 
\end{pmatrix}.
\label{eq:spurion}
\end{align}
where the elements in gray are associated with the $\Delta(B+L)=0$ operators in \cref{tab:dim6ope}.
    
By substituting the spurion matrices from \cref{eq:spurion} into \cref{Eq.BNV-ChPT} and expanding the pseudoscalar meson matrix to the zeroth order, we obtain the following  two-point BNV interactions involving an octet baryon and a lepton:
\begin{subequations}
\label{eq:ChiLBNV-Bl}
\begin{align}
 {\cal L}^{\Delta(B-L)=0}_{Bl}\supset 
& (c_1  C_{\ell uud}^{\tL\tR,x}+c_2  C_{\ell uud}^{\tL\tL,x})
(\overline{\ell_{\tL x}^{\C}}p_\tL)
-(c_1  C_{\ell uud}^{\tR\tL,x}+c_2  C_{\ell uud}^{\tR\tR,x})
(\overline{\ell_{\tR x}^{\C}}p_\tR) 
\nonumber
\\
&+(c_1  C_{\ell usu}^{\tL\tR,x}+c_2  C_{\ell usu}^{\tL\tL,x})
(\overline{\ell_{\tL x}^\C}\Sigma_\tL^+)
-(c_1  C_{\ell usu}^{\tR\tL,x}+c_2  C_{\ell usu}^{\tR\tR,x})(\overline{\ell_{\tR x}^\C}\Sigma_\tR^+)
\nonumber
\\
& + (c_1  C_{\nu dud}^{\tL\tR,x}+c_2  C_{\nu dud}^{\tL\tL,x})
(\overline{\nu_{\tL x}^{\C}}n_\tL)
+\frac{1}{\sqrt{2}}
\left[ c_1 ( C_{\nu uds}^{\tL\tR,x}- C_{\nu dsu}^{\tL\tR,x})
-c_2 C_{\nu dsu}^{\tL\tL,x} \right]
(\overline{\nu_{\tL x}^{\C}}\Sigma_\tL^0)
\nonumber
\\
&+\frac{1}{\sqrt{6}}
\left[ c_1 ( C_{\nu uds}^{\tL\tR,x}+ C_{\nu dsu}^{\tL\tR,x} -2  C_{\nu sud}^{\tL\tR,x} )
+c_2 (C_{\nu dsu}^{\tL\tL,x}-2 C_{\nu sud}^{\tL\tL,x} )\right]
(\overline{\nu _{\tL x}^{\C}}\Lambda_\tL^0) ,
\\%%
{\cal L}^{\Delta(B+L)=0}_{B l}\supset
& (c_1  C_{\bar{\ell}dds}^{\tL\tR,x}+c_2  C_{\bar{\ell}dds}^{\tL\tL,x})(\overline{\ell_{\tR x}}\Sigma_\tL^-)
-(c_1  C_{\bar{\ell}dds}^{\tR\tL,x}+c_2  C_{\bar{\ell}dds}^{\tR\tR,x})(\overline{\ell_{\tL x}}\Sigma _\tR^-)
\nonumber
\\
& -(c_1  C_{\bar{\nu}dud}^{\tR\tL,x}+c_2  C_{\bar{\nu}dud}^{\tR\tR,x})
(\overline{\nu_{\tL x}}n_\tR)
-\frac{1}{\sqrt{2}}
\left[c_1 ( C_{\bar{\nu}uds}^{\tR\tL,x}- C_{\bar{\nu }dsu}^{\tR\tL,x} )
-c_2  C_{\bar{\nu }dsu}^{\tR\tR,x} \right]
(\overline{\nu_{\tL x}}\Sigma _\tR^0)
\nonumber
\\
& -\frac{1}{\sqrt{6}}
\left[c_1 ( C_{\bar{\nu}uds}^{\tR\tL,x} + C_{\bar{\nu}{dsu}}^{\tR\tL,x} -2  C_{\bar{\nu}sud}^{\tR\tL,x})
+c_2 ( C_{\bar{\nu}dsu}^{\tR\tR,x} -2  C_{\bar{\nu}sud}^{\tR\tR,x})\right]
(\overline{\nu_{\tL x}}\Lambda_\tR^0).
\end{align}
\end{subequations}
Similarly, expansion to the first order in the meson fields gives the three-point BNV terms involving a baryon, a lepton, and a meson:
\begin{subequations}
\label{eq:ChiLBNV-BlM}
\begin{align}
{\cal L}^{\Delta(B-L)=0}_{Bl M}\supset &
\frac{i}{\sqrt{2}F_0}\Big\{
\frac{1}{\sqrt{2}}
\big[
(c_1 C_{\ell uud}^{\tL\tR,x} +c_2  C_{\ell uud}^{\tL\tL,x})
(\overline{\ell_{\tL x}^\C} p_\tL) 
+(c_1 C_{\ell uud}^{\tR\tL,x}+c_2  C_{\ell uud}^{\tR\tR,x}) 
(\overline{\ell_{\tR x}^{\C}} p_\tR )\big]\pi^0
\nonumber
\\
& - \frac{1}{\sqrt{6}}\big[
(c_1 C_{\ell uud}^{\tL\tR,x} - 3c_2 C_{\ell uud}^{\tL\tL,x})
(\overline{\ell_{\tL x}^{\C}} p_\tL) 
+(c_1 C_{\ell uud}^{\tR\tL,x} - 3c_2  C_{\ell uud}^{\tR\tR,x})
(\overline{\ell_{\tR x}^{\C}} p_\tR)\big] \eta 
\nonumber
\\
&  
+ \big[(c_1  C_{\ell usu}^{\tL\tR,x}- c_2  C_{\ell usu}^{\tL\tL,x})
(\overline{\ell_{\tL x}^{\C}} p_\tL)
+(c_1  C_{\ell usu}^{\tR\tL,x} -c_2  C_{{\ell usu}}^{\tR\tR,x})
(\overline{\ell_{\tR x}^{\C}} p_\tR)\big] \bar{K}^0
\nonumber
\\
&+(c_1 C_{\nu dud}^{\tL\tR,x}+c_2  C_{\nu dud}^{\tL\tL,x}) 
(\overline{\nu_{\tL x}^{\C}} p_\tL) \pi^-
+ \big[c_1 (C_{\nu uds}^{\tL\tR,x} +C_{\nu sud}^{\tL\tR,x} )+c_2  C_{\nu sud}^{\tL\tL,x} \big]
(\overline{\nu_{\tL x}^\C} p_\tL) K^-
\Big\}
\nonumber
\\%%
&-\frac{i}{\sqrt{2}F_0}\Big\{
\frac{1}{\sqrt{2}}(c_1  C_{\nu dud}^{\tL\tR,x}+c_2 C_{\nu dud}^{\tL\tL,x}) 
(\overline{\nu_{\tL x}^{\C}} n_\tL) \pi^0 
+\frac{1}{\sqrt{6}} 
(c_1 C_{\nu dud}^{\tL\tR,x}-3c_2  C_{\nu dud}^{\tL\tL,x})
(\overline{\nu_{\tL x}^{\C}} n_\tL) \eta 
\nonumber
\\
&
- \big[ (c_1 C_{\ell uud}^{\tL\tR,x}+c_2 C_{\ell uud}^{\tL\tL,x})
(\overline{\ell_{\tL x}^{\C}} n_\tL)
+(c_1 C_{\ell uud}^{\tR\tL,x}+c_2  C_{\ell uud}^{\tR\tR,x})
(\overline{\ell_{\tR x}^{\C}} n_\tR)\big] \pi^+ 
\nonumber
\\
&
- \big[c_1 ( C_{\nu sud}^{\tL\tR,x}+C_{\nu dsu}^{\tL\tR,x} )
+c_2( C_{\nu sud}^{\tL\tL,x}- C_{\nu dsu}^{\tL\tL,x}) \big]
(\overline{\nu_{\tL x}^{\C}} n_\tL) \bar{K}^0\Big\},
\\%%%%
{\cal L}^{\Delta(B+L)=0}_{BlM}\supset&
\frac{i}{\sqrt{2}F_0}
\Big\{
( c_1 C_{\bar{\nu}dud}^{\tR\tL,x} +c_2 C_{\bar{\nu}dud}^{\tR\tR,x}) 
(\overline{\nu_{\tL x}} p_\tR) \pi^-
+ \big[c_1( C_{\bar{\nu }uds}^{\tR\tL,x}+C_{\bar{\nu}sud}^{\tR\tL,x} )
+c_2  C_{\bar{\nu}sud}^{\tR\tR,x}\big]
(\overline{\nu_{\tL x}} p_\tR) K^- \Big\}
\nonumber
\\%%
&- \frac{i}{\sqrt{2}F_0}
\Big\{
\frac{1}{\sqrt{2}}
(c_1 C_{\bar{\nu}dud}^{\tR\tL,x}+c_2  C_{\bar{\nu}dud}^{\tR\tR,x}) 
(\overline{\nu_{\tL x}} n_\tR)\pi^0
+ \frac{1}{\sqrt{6}}
(c_1 C_{\bar{\nu}dud}^{\tR\tL,x} - 3c_2  C_{\bar{\nu}dud}^{\tR\tR,x})
(\overline{\nu_{\tL x}} n_\tR) \eta \nonumber
\\
&
-\big[
c_1 (C_{\bar{\nu }{sud}}^{\tR\tL,x}+C_{\bar{\nu }{dsu}}^{\tR\tL,x} )
+c_2 (C_{\bar{\nu }{sud}}^{\tR\tR,x}- C_{\bar{\nu }{dsu}}^{\tR\tR,x})\big]
(\overline{\nu_{\tL x}} n_\tR) \bar{K}^0
\nonumber
\\
&-\big[(c_1  C_{\bar{\ell }dds}^{\tL\tR,x}-c_2  C_{\bar{\ell}dds}^{\tL\tL,x})
(\overline{\ell_{\tR x}} n_\tL)
+( c_1 C_{\bar{\ell}dds}^{\tR\tL,x}-c_2  C_{\bar{\ell}dds}^{\tR\tR,x})
(\overline{\ell_{\tL x}} n_\tR)\big] K^- \Big\}.
\end{align}
\end{subequations}

%%%%%%%%%%%%%%%%%%%%%%%%%%%%%%%%%%%%%%
\section{The relevant BNC interactions at hadronic level}
\label{app:BNC-ChPT}
%%%%%%%%%%%%%%%%%%%%%%%%%%%%%%%%%%%%%%
First, the relevant QED vertices for the charged leptons, proton, and neutron are:
\begin{align}
\mathcal{L}^{\tt QED}_{\bar f fA}\supset
- e \bar{\ell}\gamma_\mu \ell A^\mu
+ e \bar{p}\gamma_\mu p A^\mu 
+ \frac{e\,a_p }{4m_p} \bar{p}\sigma_{\mu\nu}p F^{\mu\nu}
+ \frac{e\,a_n }{4m_n} \bar{n}\sigma_{\mu\nu}n F^{\mu\nu}.
\label{eq:qed}
\end{align} 
where $a_p=1.793$ and $a_n=-1.913$ are the anomalous magnetic moments of the proton and neutron~\cite{ParticleDataGroup:2024cfk}, respectively. 

Next, we examine the hadronic analogs of the quark-level strong and weak interactions within the framework of ChPT. 
To this end, we start with the QCD Lagrangian extended by external sources:
\begin{align}
\mathcal{L}
& =\mathcal{L}^0_{\rm QCD}
+\overline{q_\tL}\hat l^\mu \gamma_\mu q_\tL+\overline{q_\tR}\hat r^\mu \gamma_\mu q_\tR
+\left[\overline{q_\tL}(s-ip)q_\tR+\hc\right].
\end{align}
where
$\hat l_\mu,\,\hat r_\mu,\,s,\,p,\,t_l^{\mu\nu}$ are external sources associated with various quark-bilinear currents. These are $3\times3$ matrices in the flavor space and correspond to non-QCD objects. 
In our context, they can be identified as the products of leptonic currents and their coefficients shown in \cref{tab:SMweak}. 
We denote the traceless components of $\hat l_\mu$ and $\hat r_\mu$ by $l_\mu\equiv \hat l_\mu-\tr(\hat l_\mu)/3$ and $r_\mu\equiv \hat r_\mu-\tr(\hat r_\mu)/3$, respectively. 
Furthermore, we define the vector and axial-vector external sources as $\hat v_\mu \equiv (\hat l_\mu + \hat r_\mu)/2$ and $\hat a_\mu \equiv (\hat r_\mu - \hat l_\mu)/2$, respectively. Their corresponding traces are denoted by $v_\mu^s \equiv \tr(\hat v_\mu)$ and $a_\mu^s \equiv \tr(\hat a_\mu)$. 
Accordingly, we define their traceless parts as 
$v_\mu \equiv \hat v_\mu - v_\mu^s/3$ 
and $a_\mu \equiv \hat a_\mu - a_\mu^s/3$.
By incorporating these external sources, the leading order BNC chiral Lagrangian terms relevant to our analysis take the form~\cite{Jenkins:1990jv,Bijnens:1985kj,Oller:2006yh}:
\begin{align}
{\cal L}_{\tt ChPT} & 
= \frac{F^2_0}{4}\tr\left[D_\mu \Sigma (D^\mu \Sigma)^\dagger\right] 
+ \tr\left[\overline{B}  (i \slashed{D} -M_B) B\right]
\nonumber
\\
&  
+ \frac{D}{2} \tr\left[\overline{B}  \gamma^\mu \gamma_5\{u_\mu,B\}\right] 
+\frac{F}{2} \tr\left[\overline{B}  \gamma^\mu \gamma_5 [u_\mu,B]\right]
+ c_s \, a_\mu^s \tr[\overline{B} \gamma^\mu \gamma_5 B]. 
\label{eq:Bchpt}
\end{align}
Here, the covariant derivatives are defined as $D_\mu \Sigma \equiv \partial_\mu \Sigma -il_{\mu}\Sigma+i\Sigma r_{\mu}$ and $D_\mu B \equiv \partial_\mu B + [\Gamma_\mu ,B]-iv_{\mu}^s B$. $u_{\mu}$ is referred to as the chiral vielbein and is defined as 
$u_{\mu}\equiv i\left[\xi(\partial_{\mu}-ir_{\mu})\xi^{\dagger}
-\xi^{\dagger}(\partial_{\mu}-il_{\mu})\xi\right]= - i \xi^\dagger (D_\mu\Sigma )\xi^\dagger$, and $\Gamma_{\mu}$ is the chiral connection $\Gamma_{\mu}\equiv \frac{1}{2}\left[\xi(\partial_{\mu}-ir_{\mu})\xi^{\dagger}+\xi^{\dagger}(\partial_{\mu}-il_{\mu})\xi\right]$. 
$D,F,c_s$ are the LECs.
Numerically, the values $D=0.73$ and $F=0.45$ are taken from~\cite{Bali:2022qja}, while $c_s=0.16$ is determined by requiring that the contributions of the axial-vector quark currents to the nucleon matrix elements match the corresponding calculations based on the form factor method at zero momentum transfer.

By neglecting the external sources in \cref{eq:Bchpt} and expanding the meson matrix to the linear order in the meson fields, the relevant interactions between octet mesons and baryons are given by
\begin{align}
{\cal L}^{\tt QCD}_{\bar B{\tt N}M}\supset
& \frac{D + F}{2F_0} ( \overline{p} \gamma_\mu \gamma_5 p - \overline{n} \gamma_\mu \gamma_5 n ) \partial^\mu \pi^0 
+ \frac{3F - D}{2\sqrt{3}F_0} 
( \overline{p} \gamma_\mu \gamma_5 p + \overline{n }\gamma_\mu \gamma_5 n ) \partial^\mu \eta
\nonumber
\\
& + \Big(\frac{D - F}{\sqrt{2}F_0}\,\overline{\Sigma^+}\gamma_\mu \gamma_5 p 
 - \frac{D + 3F}{2\sqrt{3}F_0}\,\overline{\Lambda_0 }\gamma_\mu \gamma_5 n
 - \frac{D-F}{2F_0}\,\overline{\Sigma^0}\gamma_\mu \gamma_5 n \Big) \partial^\mu \bar{K}^0
\nonumber\\
&+ \Big( \frac{D-F}{2F_0}\,\overline{\Sigma^0}\gamma_\mu \gamma_5 p 
- \frac{D+3F}{2\sqrt{3}F_0}\,\overline{\Lambda_0 }\gamma_\mu \gamma_5 p 
+ \frac{D-F}{\sqrt{2}F_0}\,\overline{\Sigma^-} \gamma_\mu \gamma_5 n \Big) \partial^\mu K^-
\nonumber
\\
&+ \frac{D + F}{\sqrt{2}F_0}\,\overline{p} \gamma_\mu \gamma_5 n \partial^\mu \pi^+
+ \frac{D + F}{\sqrt{2}F_0}\, \overline{n} \gamma_\mu \gamma_5 p \partial^\mu \pi^- .
\label{eq:qcd}
\end{align}

Lastly, we gather the effective weak interactions that appear in the Feynman diagrams shown in \cref{tab:LO-diagrams}. 
These interactions include three-particle vertices involving a meson and a lepton current, as well as four-fermion vertices involving two baryons and two leptons. For diagrams involving neutral-current meson-lepton-lepton vertices ($M\ell\ell, M\nu\nu$), their contributions to the relevant nucleon triple-lepton decays are either suppressed (in the charged lepton case) or vanish entirely (in the neutrino case), as discussed in \cref{sec:LOdiagrams}. Therefore, we do not include these vertices here. 
For the charged-current vertices involving a meson ($M\ell\nu$), the relevant terms are 
\begin{align}
{\cal L}_{M\ell \nu}^{\tt CC} 
\supset &
 2 F_0 G_F \left[
\left(V_{ud} \partial_{\mu}\pi^- +V_{us} \partial_{\mu}K^- \right)
(\overline{\ell_{\tL x}}\gamma^{\mu}\nu_{\tL x})
+ \left(V_{ud}^{*}\partial_{\mu}\pi^+ + V_{us}^{*}\partial_{\mu}K^+\right)
(\overline{\nu_{\tL x}}\gamma^{\mu}\ell_{\tL x}) \right].
\label{eq:mesonCC}
\end{align}
The four-fermion weak interactions between the nucleon and leptonic currents are obtained by bringing the weak external sources listed in \cref{tab:SMweak} into the baryonic chiral Lagrangian terms in \cref{eq:Bchpt}, yielding the following neutral- and charged-current interactions:  
\begin{align}
{\cal L}_{{\tt NN}ll}^{\tt NC} 
\supset &
-\frac{\sqrt{2}G_F}{6}
\big[\bar{p}\gamma^{\mu}
(3(1-4s_W^2) - (2D+6F-3 c_s)\gamma_5)p
+\bar{n}\gamma_{\mu}(-3 + (4D+3 c_s)\gamma_5) n\big] 
\nonumber
\\
&
\times \big[
(\overline{\nu_{\tL x}}\gamma_{\mu}\nu_{\tL x})+(-1+2s^2_W)(\overline{\ell_{\tL x}}\gamma_{\mu}\ell_{\tL x})+2s^2_W(\overline{\ell_{\tR x}}\gamma_{\mu}\ell_{\tR x})
\big],
\label{eq:baryonNC}
\\%%charged current
{\cal L}_{B{\tt N}\ell \nu}^{\tt CC} \supset &
-\sqrt{2}G_F \Big[ V_{ud}^{*}\, \overline{n}\gamma_{\mu}(1-(D+F)\gamma_5) p
-\frac{V_{us}^{*}}{\sqrt{6}}\, \overline{\Lambda^0}\gamma_{\mu}(3 - (D+3 F)\gamma_5) p
\nonumber
\\
&
-\frac{V_{us}^{*}}{\sqrt{2}}\,\overline{\Sigma^0}\gamma^{\mu}(1 +(D-F)\gamma_5) p
- V_{us}^{*}\, \overline{\Sigma^-}\gamma_{\mu}(1+(D-F)\gamma_5)n
\Big] (\overline{\nu_{\tL x}}\gamma^{\mu}\ell_{\tL x})
\nonumber
\\
&
-\sqrt{2}G_F \Big[ V_{ud}\, \overline{p}\gamma_\mu(1 - (D+F)\gamma_5)n \Big]
(\overline{\ell_{\tL x}}\gamma^{\mu}\nu_{\tL x}).
\label{eq:baryonCC}
\end{align}    

%%%%%%%%%%%%%%%%%%%%%%%%%%%%%%%%%%%%%%
\section{Experimental limits on two-body nucleon decay modes and branching ratios of octet meson leptonic decays}
\label{app:meson_decay}
%%%%%%%%%%%%%%%%%%%%%%%%%%%%%%%%%%%%%%

\cref{tab:data} summarizes the current experimental limits on two-body nucleon decays involving a lepton and a pseudoscalar meson, as well as the leptonic decay branching ratios of pseudoscalar and vector mesons. 
These data are used to derive constraints on the partial lifetimes of the corresponding nucleon triple-lepton decay modes due to the meson-mediated diagrams shown in \cref{tab:LO-diagrams}. 
Note that the branching ratios for the pseudoscalar meson $K_S^0$ and  the vector mesons $\rho^\pm,K^{*\pm}$  are based on theoretical estimates from \cite{Ecker:1991ru,Yang:2021crs}, while the remaining branching ratios correspond to central values or the 90\% CL upper bounds (indicated by ``$<$") compiled by the latest PDG~\cite{ParticleDataGroup:2024cfk}.

\begin{table}[t]
\centering
\resizebox{0.85\linewidth}{!}{
\renewcommand{\arraystretch}{1.1}
\begin{tabular}{|c|c|c|c|c|c|c|c|}
\hline
 \multicolumn{2}{|c|}{Proton decay}  
& \multicolumn{2}{c|}{Neutron decay} 
& \multicolumn{2}{c|}{Pseudoscalar meson} 
& \multicolumn{2}{c|}{Vector meson} 
\\
\hline
\multicolumn{1}{|c|}{Mode} 
& \multicolumn{1}{c|}{$\Gamma_{\tt exp}^{-1}\,[\rm yr]$}  
& \multicolumn{1}{c|}{Mode} 
& \multicolumn{1}{c|}{$\Gamma_{\tt exp}^{-1}\,[\rm yr]$}  
& \multicolumn{1}{c|}{Mode} 
& \multicolumn{1}{c|}{Branching ratio} 
& \multicolumn{1}{c|}{Mode} 
& \multicolumn{1}{c|}{Branching ratio} 
\\
\hline%%
\multirow{2}{*}{$p\to\nu\pi^+$} & \multirow{2}{*}{$3.9\times 10^{32}$} 
& $n\to e^+\pi^-$ & $5.3\times 10^{33}$  
& $\pi^{\pm}\to e^{\pm}\nu_e$ & $1.23\times 10^{-4} $ 
& $\rho^\pm \to e^\pm \nu_e $ & $4.6\times 10^{-13}$~\cite{Yang:2021crs} 
\\ 
& 
& $n\to \mu^+\pi^-$  & $3.5\times 10^{33}$ 
& $\pi^{\pm}\to \mu^{\pm}\nu_\mu$ & $0.999877$
& $\rho^\pm \to \mu^\pm \nu_\mu $ & $4.5\times 10^{-13}$~\cite{Yang:2021crs} 
\\ 
\hline%%
\multirow{2}{*}{$p\to \nu K^+$} & \multirow{2}{*}{$5.9\times 10^{33}$}
& $n\to e^-K^+$  & $3.2\times 10^{31}$ 
& $K^{\pm}\to e^{\pm}\nu_e$ & $1.582\times 10^{-5}$ 
& $K^{*\pm} \to e^\pm \nu_e $ & $1.2\times 10^{-13}$~\cite{Yang:2021crs}
\\ 
& 
& $~n\to \mu^-K^+~$  & $5.7\times 10^{31}$ 
& $K^{\pm}\to \mu^{\pm}\nu_\mu$ & $0.6356$ 
& $~K^{*\pm} \to \mu^\pm \nu_\mu~$ & $~1.2\times 10^{-13}$~\cite{Yang:2021crs}~
\\ 
\hline%%
$p\to e^+ \pi^0$ & $2.4\times10^{34}$ 
& &  
& $\pi^0\to e^-e^+$ & $6.46\times 10^{-8}$
& $\rho^0\to e^+e^-$ & $4.72\times 10^{-5}$
\\ 
$p\to e^+ \eta $& $10^{34}$ 
& $n\to \nu \pi^0$  & $1.1\times10^{33}$
& $\eta\to e^-e^+$ & $<7\times 10^{-7}$
& $\omega\to e^+e^-$ & $7.41\times 10^{-5}$   
\\
$p\to \mu^+ \pi^0$ & $1.6\times10^{34}$ 
& $n\to \nu \eta$ & $~1.58\times10^{32}~$   
& \multirow{2}{*}{$\eta\to \mu^-\mu^+$} & \multirow{2}{*}{$5.8\times 10^{-6}$} 
& $\rho^0\to \mu^+\mu^-$ & $4.55\times 10^{-5}$
\\ 
$p\to \mu^+\eta$& $4.7\times10^{33}$&  &   
& &
& $\omega\to \mu^+\mu^-$ & $7.4\times 10^{-5}$
\\ 
\hline%%
$p \to e^+ K_L^0$& $5.1\times10^{31}$  
& &  
& $K_L^0\to e^+ e^-$  & $(9^{+6}_{-4})\times 10^{-12}$
& $K^{*0}\to e^+ e^-$ &  ---
\\ 
$p \to e^+ K_S^0$& $1.2\times10^{32}$   
& $n\to\nu K_L^0$ & ---
& $K_S^0\to e^+ e^-$ & $2.1\times10^{-14}$~\cite{Ecker:1991ru}
& $\bar K^{*0}\to e^+ e^-$ &  ---
\\ 
$p \to \mu^+ K_L^0$& $8.3\times10^{31}$  
& $n\to\nu K_S^0$ & $2.6\times10^{32}$
& $K_L^0\to \mu^+ \mu^-$  & $6.84\times 10^{-9}$
& $K^{*0}\to \mu^+ \mu^-$ &  --- 
\\ 
$~p \to \mu^+ K_S^0~$&$~1.5\times10^{32}~$
& &  
&$~K_S^0\to \mu^+ \mu^-~$ & $~5.1\times 10^{-12}$~\cite{Ecker:1991ru}~ 
& $\bar K^{*0}\to \mu^+ \mu^-$ &  --- 
\\
\hline
\end{tabular}}
\caption{Summary of the experimental lower bounds on the partial lifetimes of two-body nucleon decays and the branching ratios of meson decays into two leptons. 
}
\label{tab:data}
\end{table}

\twocolumngrid
\bibliography{references_paper}{}
\bibliographystyle{utphys}
%%%%%%%%%%%%%%%%%%%%%%%%
\end{document}